\documentclass[reprint,
  aps,
  prd,
  superscriptaddress,
  nofootinbib
]{revtex4-2}
\usepackage[english]{babel}
\usepackage{multirow}
\usepackage{graphicx}
\usepackage{dcolumn}
\usepackage{amsmath}
\usepackage{amssymb}
\usepackage{hyperref}
\usepackage{xcolor}
\usepackage{subcaption}
\usepackage{enumitem}
\usepackage{soul}
\usepackage[normalem]{ulem}
\hypersetup{    
  colorlinks      = true,
  linkcolor       = {blue},
  linkbordercolor = {white},
  citecolor       = {blue},
  citebordercolor = {white},
  urlcolor        = {blue},
  urlbordercolor  = {white},
}
\usepackage{bm}
\usepackage{color}

\newcommand{\ben}{\begin{enumerate}}
\newcommand{\een}{\end{enumerate}}
\newcommand{\bea}{\begin{eqnarray}}
\newcommand{\eea}{\end{eqnarray}}
\newcommand{\be}{\begin{equation}}
\newcommand{\ee}{\end{equation}}

\def\rm{\rm}

\usepackage{scrextend}
\usepackage{comment}

\usepackage{tikz}
\usetikzlibrary{positioning, shapes.geometric, arrows.meta}

\def\nn{\nonumber}
\def\pa{\partial}
\def\l{\left}
\def\r{\right}

\def\l{\left}
\def\r{\right}
\def\a{\alpha}
\def\b{\beta}
\def\c{\gamma}
\def\d{\delta}
\def\m{\mu}
\def\n{\nu}
\def\w{\omega}

\def\La{\Lambda}

\def\O{\Omega}

\def\bs{\boldsymbol}
\def\eps{\epsilon}

\def\rm{\mathrm}

\def\*{\star}

\newcommand{\ie}{\text{i.e.,~}}

\begin{document}

\renewcommand{\arraystretch}{1.5} 
\setlength{\tabcolsep}{0.2cm} 

\title{Forecasting Constraints on Non-Thermal Light Massive Relics from Future CMB Experiments (CMB-S4/Simons Observatory)}

\author{Arka Banerjee}
\affiliation{Department of Physics, Indian Institute of Science Education and Research, Homi Bhabha Road, Pashan, Pune 411008, India}

\author{Abhik Bhattacharjee}
\email{abhikbhattacharjee@hri.res.in}
\address{Harish-Chandra Research Institute, A CI of Homi Bhabha National Institute, Chhatnag Road, Jhunsi, Prayagraj, Uttar Pradesh 211019, India}

\author{Subinoy Das}
\address{Indian Institute of Astrophysics, Bengaluru, Karnataka 560034, India}

\author{Anshuman Maharana}
\address{Harish-Chandra Research Institute, A CI of Homi Bhabha National Institute, Chhatnag Road, Jhunsi, Prayagraj, Uttar Pradesh 211019, India}


\author{Ravi Kumar Sharma}
\email{rksharma@physik.rwth-aachen.de}

\affiliation{Institute for Theoretical Particle Physics and Cosmology  (TTK),
RWTH Aachen University, Sommerfeldstr. 16, D-52056 Aachen, Germany}

\begin{abstract}

Precise measurements of the cosmological impact of dark sector relics can shed light on Physics beyond the Standard Model. In this work we present Fisher forecasts on \textit{non-thermal LiMR} models for {\color{black} a CMB Stage IV-like experiment and the Simons Observatory}---particularly focusing on a model of inflaton/moduli decay giving rise to non-thermally distributed dark sector particles, and also comparing our results with those for sterile particles following the Dodelson-Widrow distribution. Two independent parameters, the effective number of extra relativistic species $\Delta N_\rm{eff}$ and the effective mass $M_\rm{sp}^\rm{eff}$ of the relic, influence linear cosmological observables. We find $\Delta N_\rm{eff}$ to be more tightly constrained with $\sigma(\Delta N_\rm{eff})\sim10^{-3}$, for a less abundant, heavier LiMR which becomes fully non-relativistic around matter-radiation equality than a more abundant, lighter LiMR which becomes fully non-relativistic just after recombination, for which $\sigma(\Delta N_\rm{eff})\sim10^{-2}$. The uncertainties on $M_\rm{sp}^\rm{eff}$ differ by a factor of $\sim3$ between the two cases. Our analysis also reveals distinct parameter correlations: the phenomenological parameters $\{\Delta N_\rm{eff},M_\rm{sp}^\rm{eff}\}$ are found to be negatively correlated for the former case and positively correlated for the latter. We obtain similar {\color{black}projected uncertainties} on the cosmological parameters (in either case) for both the inflaton/moduli decay and the Dodelson-Widrow models when the first two moments of the LiMR distribution function, related to the phenomenological parameters, are matched. Finally, by constructing a modified distribution that matches the first two moments of the Dodelson-Widrow but deviates maximally in the third moment, we demonstrate that CMB Stage IV data is not expected to be sensitive to higher moments of the distribution.
\end{abstract}


\maketitle

\section{Introduction} \label{sec:intro}
The Cosmic Microwave Background (CMB) provides a snapshot of the Universe at the time of recombination -- when electrons and protons combined to form hydrogen atoms -- which occurred about 380, 000 years after the end of inflation. Post recombination, photons decoupled from the cosmic plasma giving rise to the CMB. Likewise, neutrinos decoupled from the plasma when the expansion rate of the universe exceeded the rate of weak interactions, when the Universe was about a second old. This gave rise to the Cosmic Neutrino Background (C$\nu$B), which has not been detected yet, but analysis of CMB anisotropies~\cite[e.g.][]{Hou_2013, Follin_2015, Baumann2016} and agreement between calculated and observed primordial abundances of light elements indirectly establishes its presence~\cite{Lesgourgues_2012}. An important observable in any cosmological model is the energy density in radiation at the time of recombination. Within the Standard Model (SM), photons and neutrinos are the sole contributors to this. More generally, with Beyond the Standard Model (BSM) physics, there can be contributions from other species. Observational results are usually reported by making use of the quantity $N_\rm{eff}$, the effective number of neutrino like species contributing to the radiation energy density before photon decoupling, in addition to that of photons. Current constraints on $N_\rm{eff}$ from CMB measurements are $2.99\pm0.17$ ($68\%$ CL; Planck TT,TE,EE+lensing+BAO)~\cite{Planck:2018vyg} and $N_\rm{eff}<3.08$ (one tail $95\%$; P-ACT-LB)~\cite{calabrese2025atacamacosmologytelescopedr6}, which are consistent with the SM prediction of $N_\rm{eff}=3.044$~\cite{Akita_2020, Froustey_2020, Bennett_2021}. However, there is still room for $\Delta N_\rm{eff}\equiv N_\rm{eff}-3.044\neq0$ once we include recent BAO  measurements from DESI DR2: $N_\rm{eff}=3.10\pm0.17$ ($68\%$ CL; DESI+CMB)~\cite{osti_2346175} and $N_\rm{eff}=3.23\pm0.18$ ($68\%$ CL; DESI BAO+CMB)~\cite{elbers2025constraintsneutrinophysicsdesi}. So the possibility of the existence of extra relativistic particles is not ruled out yet. Furthermore, $N_{\rm {eff}}$ will be probed with even greater precision with future CMB experiments. {\color{black}For example, an experiment with the specifications of the CMB Stage IV (CMB-S4) survey will have far better sensitivity to $N_\rm{eff}$, with a target of $\sigma(N_\rm{eff})=0.02-0.03$ for the $\La$CDM model extended by $N_\rm{eff}$~\cite{abazajian2016cmbs4sciencebookedition}}.
A quantitative understanding of possible constraints on theoretical models with the expected data quality from CMB-S4 is timely.

The presence of new light relics which behave as ``dark radiation" and contribute to $\Delta N_\rm{eff}$, is a generic feature of many BSM proposals. Sources for dark radiation include sterile neutrinos~\cite[e.g.][]{Abazajian_2001, Boyarsky_2009, abazajian2012lightsterileneutrinoswhite} and axions~\cite[e.g.][]{Grin_2008,Brust_2013,Salvio_2014,Kawasaki_2015,Baumann_2016,Hlo_ek_2017,visinelli2025axionsdarkmatterdark}, among others\footnote{See Ref.~\cite{abazajian2016cmbs4sciencebookedition} (and references therein) for discussion of other possibilities and their prospects for detection by the CMB-S4 experiment.}. Certain studies propose the existence of sterile neutrinos~\cite{PhysRevD.84.073008,conrad2012sterileneutrinofitsshort,Gariazzo_2017} to address anomalies in neutrino oscillation data~\cite{Aguilar_2001,Aguilar_Arevalo_2018}. In spite of the particle physics motivation, a fully thermalized fourth sterile neutrino which decoupled along with the standard neutrinos is ruled out due to stringent constraints on $N_\rm{eff}$ mentioned above~\cite{Planck:2018vyg}. Models with more complicated structures for the neutrino sector have also been proposed \cite[e.g.][]{Arkani-Hamed:2016rle}, and possible constraints from cosmological data have been investigated~\cite{Banerjee:2016suz}. However, hot thermal relics which decoupled from the SM plasma at an earlier time than neutrino decoupling could be exempted from the strong constraints on $N_\rm{eff}$. Thermal relics 
with decoupling temperatures above the top quark mass have the following minimal contributions: $\Delta N_\rm{eff}=0.027$ for a Goldstone boson, $\Delta N_\rm{eff}=0.047$ for a Weyl fermion, and $\Delta N_\rm{eff}=0.054$ for vector boson~\cite{https://doi.org/10.17863/cam.30368}. The cosmological impact of such massless light relics (or relics with masses $m\ll\rm{eV}$) is completely captured by the $N_\rm{eff}$ parameter~\cite{Brust_2013}. 
They primarily affect the expansion of the Universe, causing a suppression of the damping tail of the CMB power spectrum as well as a phase shift in the Baryon Acoustic Oscillations~\cite{Bashinsky_2004, Hou_2013, Brust_2013, Baumann_2016, green2019messengersearlyuniversecosmic}. 

The other possibility is that of Light Massive Relics (LiMRs) ---  massive particles which contributed to the radiation energy density in the early Universe and to the matter energy density at late times. The LiMR mass is an important parameter as it controls the time at which the LiMR transitions to being non-relativistic, after which it behaves as a sub-component of the total Dark Matter (DM) component of the Universe. Increasing the effective mass of the LiMR primarily affects the CMB power spectrum at low multipoles through the late Integrated Sachs-Wolfe (ISW) effect and CMB lensing~\cite{Lesgourgues_2012,Pan_2015, Lesgourgues_Mangano_Miele_Pastor_2013}.  In Ref.~\cite{Xu_2022}, the authors combined CMB and LSS data to obtain limits on the masses of thermally decoupled species $X$ (with a present day temperature of $T_X^{(0)}=0.91~\rm K$): $m_X\leq 11,2.3,1.6~\rm{eV}$ for scalars, Weyl fermions, and vectors respectively. Fisher forecasts for such LiMRs, with masses ranging from $10^{-2}~\rm{eV}\leq m_X\leq 10~\rm{eV}$ and temperatures in the range $[0.91~\rm{K},1.50~\rm{K}]$, have been performed in Ref.~\cite{DePorzio_2021} to {\color{black}predict} constraints on $g_X$, the number of degrees of freedom of the LiMR, from the CMB-S4 experiment.

While much of the literature has focused on thermal relics, this paper deals with  LiMRs in  \textit{non-thermal} distributions. Various well-motivated models have been proposed to generate species with non-thermal distributions in the early Universe. One of the familiar categories is through mixing of active neutrinos~\cite{Dodelson_1994,PhysRevLett.82.2832}, while another is through the decay of heavy particles in the early Universe~\cite{Cuoco_2005,1304.1804,Hasenkamp_2013,Acharya_2019, Bhattacharya:2020zap, Baumholzer:2021heu}. Belonging to the latter category is the perturbative decay of the inflaton~\cite{Mukhanov_2005} or the decay of the moduli due to vacuum misalignment (see, for instance, Ref.~\cite{Cicoli_2016}). 
We focus on two concrete examples of such distributions to draw out the main features: the first arises from the inflaton/moduli decay in the early Universe~\cite[e.g.][]{1304.1804, Bhattacharya:2020zap}. The second distribution we examine is the Dodelson-Widrow distribution~\cite{Dodelson_1994,Acero_2009}, originally proposed for sterile neutrinos. We perform Fisher forecasts for these models to quantify the {\color{black}expected uncertainties} on their parameters that are expected from a survey with the specifications of the CMB-S4 experiment\footnote{{\color{black}The preparation of this manuscript was completed before the descoping of the CMB-S4 project. Since then, we have revised the manuscript and included forecasts for a cosmic variance–limited experiment in addition to those of a CMB-S4–like experiment. We have also included forecasts for the Simons Observatory-Large Aperture Telescope (SO-LAT) in Section~\ref{sec:results}.}}. We also investigate how features of the shape of the LiMR distribution function can affect observables.

The paper is structured as follows: Section \ref{sec:Review} gives a brief
review of some basic aspects of the LiMR models, Fisher forecasts  and future experimental prospect.  We discuss our implementation in section \ref{sec:imp}. Our results are presented in Section \ref{sec:results}. In Section \ref{sec:higher}, we discuss the effect of the shape of the distribution function. We conclude in Section \ref{sec:conclusion}.

\section{Review} \label{sec:Review}

\subsection{Light Massive relics} \label{sec:LiMR}

We begin by discussing some general aspects of LiMR models. The influence of LiMRs on linear cosmological observables can be characterized by three parameters (see Ref.~\cite{Acero_2009} for details).

\begin{itemize}
    \item $\Delta N_\rm{eff}$: This parameter represents the contribution of  LiMRs to the Universe's  relativistic energy density before photon decoupling. It is defined as ($g_s$ is a  degeneracy factor):
    \begin{equation}\label{1}
        \Delta N_\rm{eff}\equiv\frac{\rho^\rm{rel}_\rm{sp}}{\rho_\nu}=\frac{g_s}{7/4}\l[\frac{1}{\pi^2}\int dp~p^3\hat f(p)\r]/\l[\frac{7\pi^2}{120}(T_\n^\rm{id})^4\r],  
    \end{equation}
    where $\hat f(p)$ is the momentum distribution function of the  {LiMR},  and $T_\nu^\rm{id} \equiv (4/11)^{1/3} T_{\gamma}$ is the neutrino temperature in terms of the photon temperature, assuming instantaneous neutrino decoupling. 
    
    \item $M_\rm{sp}^\rm{eff}$: This parameter represents the contribution of the LiMR to the Universe's current energy density, and it is defined as ($g_s'$ is a degeneracy factor):
    \begin{align}\label{2}
        \frac{M_\rm{sp}^\rm{eff}}{94.05~\rm{eV}}\equiv \omega_\rm{sp}= \frac{g_s'}{3/2}\left[\frac{m_\rm{sp}}{\pi^2}\int dp~p^2\hat f(p)\right]\left[\frac{h^2}{\rho_\rm{crit}^0}\right],
    \end{align}
    where $h$ is the reduced Hubble constant and $\rho^0_\rm{crit}$ denotes the critical density of the Universe today.
    
    \item $\left<v_\rm{fs}\right>$: The typical free-streaming velocity of the particles today, given by
    \begin{equation}\label{3}
        \left<v_\rm{fs}\right>\equiv\frac{\frac{g_s}{7/4}\int dp~p^2\frac{p}{m_\rm{sp}}\hat f(p)}{\frac{g_s'}{3/2}\int dp~p^2\hat f(p)} = 5.236 \times 10^{-4}\frac{\Delta N_\rm{eff}}{M_\rm{sp}^\rm{eff}}.
    \end{equation}
\end{itemize}

Note that only two of the three parameters are independent. In our study, we will utilize $M_\rm{sp}^{\rm{eff}}$ and $\Delta N_\rm{eff}$ as phenomenological parameters to investigate the impact of LiMRs. 
Also, the above described parameters are determined by the first two moments of the distribution function. We will examine the role of higher moments in section \ref{sec:higher}. Note that the parameter $g_X$ of Ref.~\cite{Xu_2022} is related to our $g_s,g_s'$. But unlike in Ref.~\cite{Xu_2022}, where the {\color{black}uncertainties} on $\Delta N_\rm{eff}$ or $M_\rm{sp}^\rm{eff}$ (equivalently $\O_X$) were obtained by translating the {\color{black}predicted uncertainty} on $g_X$, here we shall find the uncertainties on these phenomenological parameters by varying the mass $m_\rm{sp}$ and a model specific parameter (see below) which enters into the momentum distribution function.

Next, we describe the two LiMR distribution functions that we will consider in the paper.

\subsubsection{Distribution function associated with production via decay of the Inflaton/Moduli}
During reheating after inflation or an epoch of moduli domination, the early Universe can go through a matter-dominated era with its energy density dominated  by heavy cold particles ($\phi$). In addition to the
decay to the SM sector, the $\phi$ particle can decay to a LiMR in the dark sector. This typically occurs via a $1 \to 2$ process:
\begin{align*}
    \phi\xrightarrow{{B_\rm{sp}}} \texttt x\texttt x,
\end{align*}
where \texttt x stands for the LiMR\footnote{{\color{black} Here $\mathbf x$ behaves as a hot/warm dark matter candidate. Cold dark matter production in an early matter-dominated era (EMDE) has also been studied, for instance, in Refs.~\cite{Drees_2018,Allahverdi_2019}. Several other works have also examined how an EMDE modifies dark-matter production, cosmological evolution, and particle constraints, including studies of ultra-cold WIMPs, gravitational heating, freeze-in and freeze-out during EMDEs, 21-cm signatures, and related non-standard thermal histories~\cite{Miller:2019pss, Chung_1999, gelmini2010dmproductionmechanisms, Gelmini_2008, Ganjoo_2024, Bae_2025, Allahverdi_2022, Dutra_2023, Silva_Malpartida_2025, PhysRevD.108.043533, Banerjee_2022, Ghoshal_2022, Ling__2025}. Also see Ref.~\cite{arcadi2024thermalnonthermaldmproduction} for a mini-review of thermal and non-thermal DM production in non-standard cosmologies.}}. The production rate of \texttt x is controlled by the decay lifetime $\tau$, the mass $m_{\phi}$, and the branching ratio $B_\rm{sp}$. 
LiMRs produced via this mechanism have the following momentum  distribution (as given in Ref.~\cite{Bhattacharya:2020zap}; see Appendix~\ref{appendix:A} for further details):
\begin{align}\label{fqNT}
    f(\bs q)=\frac{32}{\pi\hat E^3}\l(\frac{N(0)B_\rm{sp}}{\hat s^3(\theta^*)}\r)\frac{\exp{(-\hat s^{-1}(y))}}{|\bs q|^3\hat H(\hat s^{-1}(y))}.
\end{align}
Here $\hat E$ is the energy of the particles at production, $N(0)$ is the initial number density of the $\phi$ particles, $\hat s(\theta)\equiv a(t)$ is the scale factor as a function of dimensionless time $\theta$ (with $\theta^*$ indicating the time by which all the $\phi$ particles have decayed), $\hat s^{-1}$ is the functional inverse of the scale factor function, and $\hat H=\hat s'(\theta)/\hat s(\theta)$ is the dimensionless Hubble parameter. {\color{black} The argument of the distribution function is $\bs q\equiv\bs p/T_\rm{ncdm,0}$ where $T_\rm{ncdm,0}$ is the typical momentum of the LiMRs today. Motivated by the expectation that most $\phi$ particles decay well before $\theta^*$, the authors in Ref.~\cite{Bhattacharya:2020zap} took $T_\rm{ncdm,0}=\frac{\hat E}{4}\frac{a(t^*)}{a(t_0)}$ which leads to the following constraint on $|\bs q|$, when coupled with Eq. (\ref{A8}),}
\begin{align}
     \frac{4}{\hat s(\theta^*)}<|\bs q|<4.
\end{align}

The model has four microscopic parameters: $m_\phi,\tau, B_\rm{sp}$ and $m_\rm{sp}$. We  will take the first two parameters to be $m_\phi\sim10^{-6}M_\rm{Pl}$ and $\tau\sim10^8/m_\phi$  (motivated by having $\phi$ to be driving inflation at the GUT scale and decaying by GUT scale interactions). The phenomenological parameters $\{\Delta N_\rm{eff},M_\rm{sp}^\rm{eff}\}$ are related to the model parameters by the following expressions~\cite{Bhattacharya:2020zap}:
\begin{align}
    \Delta N_\rm{eff}&=\frac{43}{7}\frac{B_\rm{sp}}{1-B_\rm{sp}}\l(\frac{g_*(T(t_\n))}{g_*(T(t^*))}\r)^{1/3},\label{DNeff}\\
    M_\rm{sp}^\rm{eff}&=\frac{62.1m_\rm{sp}}{g^{1/4}_*(T(t^*))}\frac{B_\rm{sp}}{(1-B_\rm{sp})^{3/4}}\l(\frac{M_\rm{Pl}}{\tau m_\phi^2}\r)^{1/2}\label{Meff},
\end{align}
where $t_\n$ is the neutrino-decoupling time and $g_*(T(t))$ is the effective number of degrees of freedom at a temperature $T$ (corresponding to a time $t$).

\subsubsection{The Dodelson-Widrow Distribution}
We also consider the  Dodelson-Widrow distribution~\cite{Dodelson_1994}. Non-resonant active-sterile neutrino oscillations in the early Universe in the limit of small mixing angle and zero leptonic asymmetry leads to the following distribution function for the sterile neutrinos~\cite{Acero_2009}:
\begin{align}
    f_\n(p)=\frac{\chi}{e^{p/T_\n}+1},
\end{align}
where $\chi$ is an arbitrary normalization factor and $T_\n$ is the temperature of neutrinos today. With $T_\n=T_\n^\rm{id}$,
the phenomenological parameters are
related to the model parameters via the following relations:
\begin{align}\label{DWparams}
    \Delta N_\rm{eff}=\chi\quad\text{and}\quad M_\rm{sp}^\rm{eff}=m_\rm{sp}\times\chi,
\end{align}
where $m_\rm{sp}$ is the mass of the sterile neutrinos. 

\subsection{Fisher Forecast Methodology}\label{sec:Methodology}
\subsubsection{Formalism}
Suppose we have a set of data $\{\bs d\}$ and a model $\mathcal M$ with a set of parameters $\{\bs\theta\}$. We want to find the probability distribution of $\{\bs\theta\}$, given the data $\{\bs d\}$, \ie $P(\bs\theta|\bs d)$, which is obtained using Bayes' theorem:
\begin{align}
    P(\bs\theta|\bs d)\propto P(\bs d|\bs\theta)P(\bs\theta)
\end{align}
Here $P(\bs\theta)$ and $P(\bs\theta|\bs d)$ are called the \textit{prior} and \textit{posterior} distributions respectively and $P(\bs d|\bs\theta)$ is called the \textit{Likelihood}, usually denoted by $\mathcal{L}$, and can be expressed as follows~\cite{Durrer_2020}:
\begin{align}\label{like}
    \mathcal{L}(\bs d;\bs \theta)=\frac{1}{\sqrt{(2\pi)^n|\mathbf C(\bs\theta)|}}\exp{\l(-\frac{1}{2}\bs d^T\mathbf C(\bs\theta)\bs d\r)},
\end{align}
where $\mathbf{C}$ represents the covariance matrix {\color{black}and $n$ is the dimension of the data vector $\bs d$}. The Fisher matrix is related to the curvature of the log-likelihood function, evaluated at the fiducial values of the parameters:
\begin{equation}
    F_{ij} = -\l<\frac{\partial^2 \ln{\mathcal L}}{\partial \theta_i \partial \theta_j}\r>\bigg\rvert_{\bs{\theta=\theta_0}},
\end{equation}
$\theta_i,\theta_j$ being the parameters of interest and $ \bs\theta_0$ being the fiducial set of parameters~\cite{Tegmark_1997}.\\
{\textit{Derived Parameters:}}
If we have a Fisher matrix in terms of a set of parameters $\{\bs\theta\}$, we can obtain a new Fisher matrix in terms of a set of derived parameters $\{\bs\theta'\}$ as~\cite{coe2009fishermatricesconfidenceellipses}
\begin{align}\label{jac}
    F'=J^T\cdot F\cdot J,
\end{align}
where $J$ stands for the inverse Jacobian matrix $\pa\bs\theta/\pa\bs\theta'$. This will be used, for instance, to translate the $1\sigma$ uncertainties on $\{B_\rm{sp},m_\rm{sp}\}$ to those on $\{\Delta N_\rm{eff}, M_\rm{sp}^\rm{eff}\}$.\\
{\textit{Error Estimate:}}
After computing the Fisher information matrix, we can derive the error covariance matrix by taking its inverse
\begin{equation}
\rm{Cov}_{ij} = (F^{-1})_{ij}. 
\end{equation}
The diagonal elements of the error covariance matrix indicate the marginalized errors on the parameters. For instance, the anticipated marginalized $1\sigma$ error on a parameter $\theta_i$, accounting for all degeneracies concerning other parameters, is calculated as:

\begin{equation}\label{sigma_i}
\sigma_i^{(\rm{marg.})} = \sqrt{\rm{Cov}_{ii}}.
\end{equation}

The unmarginalized expected errors, or conditional errors, can be determined by
\begin{equation}
\sigma_i^{(\rm{cond.})} = \sqrt{\frac{1}{F_{ii}}},
\end{equation}
representing the square root of the reciprocal of the relevant diagonal element of the Fisher matrix.\newline
\textit{Posterior Probability or confidence regions:}
For the form of the Likelihood function (\ref{like}), the posterior probability is Gaussian  (under the Fisher assumption\footnote{One can have non-Gaussian posteriors in general even for a Gaussian likelihood when priors are non-Gaussian.}) with the $1\sigma$ and $2\sigma$ intervals encompassing $68.3\%$ and $95.4\%$ of the total probability respectively. For Gaussian posterior, the confidence regions are ellipses in the parameter space, whose size and orientation are determined by the Fisher matrix elements. Confidence regions correspond to regions in parameter space in which the joint parameter values are expected to lie with a certain probability -- $1\sigma$ region corresponds to $39.35\%$ probability and $2\sigma$ region corresponds to $86.47\%$ probability. 
\\
\subsubsection{CMB Experiments}
 
In the context of Cosmic Microwave Background (CMB), the data vector is either the CMB power spectra or spherical harmonic coefficients represented as $\textbf{d} = \{a_{\ell m}^{T}, a_{\ell m}^{E}, a_{\ell m}^{\phi}\}$. The Fisher information matrix takes the following form~\cite{Wu_2014}:
\begin{equation}\label{Fij}
    F_{ij} = \sum_\ell \frac{2\ell + 1}{2} f_{\text{sky}} \text{Tr}\left(\mathbf{C}_\ell^{-1}(\boldsymbol{\theta}) \frac{\partial \mathbf{C}_\ell}{\partial \theta_i} \mathbf{C}_\ell^{-1}(\boldsymbol{\theta}) \frac{\partial \mathbf{C}_\ell}{\partial \theta_j}\right),
\end{equation}
where $f_\rm{sky}$ is the fractional sky coverage, $\bs\theta$ represents the vector of parameters, and $\mathbf{C}_\ell$ is the total covariance matrix of the relevant CMB observables (temperature, E-mode polarization, and lensing potential) given by~\cite{osti_1244517}:
\[
\mathbf{C}_\ell = \begin{pmatrix}
    C_{\ell}^{TT}+N_{\ell}^{TT} & C_{\ell}^{TE} & C_{\ell}^{T\phi} \\
    C_{\ell}^{TE} & C_{\ell}^{EE}+N_{\ell}^{EE} & 0 \\
    C_{\ell}^{T\phi} & 0 & C_{\ell}^{\phi\phi}+N_{\ell}^{\phi\phi} \\
\end{pmatrix}
\]
This includes the Gaussian noise $N_\ell^{XX}$, where $XX\in\{TT,EE\}$, given by~\cite{abazajian2016cmbs4sciencebookedition}
\begin{align}\label{27}
    N_\ell^{XX}=s^2\exp{\l(\ell(\ell+1)\frac{\theta^2_\rm{FWHM}}{8\ln{2}}\r)},
\end{align}
where $\theta_\rm{FWHM}$ is the resolution of the experiment in arcmin, and $s$ is the instrumental noise in temperature/polarization.\\

\section{Implementation}
\label{sec:imp}
\begin{table*}
    \centering
    \begin{tabular}{|c|c|c|c|c|}
    \hline
      Experiment  &  $\ell$ range & Noise $s [\rm{\m K-arcmin}]$ & $f_\rm{sky}$ & $\theta_\rm{FWHM}[\rm{arcmin}]$\\\hline\hline
      Planck  & T: $2-2500$ & T: $[145,149,137,65,43,66,200]$ & $0.6$ & $[33,23,14,10,7,5,5]$\\
              & P: $30-2500$ & P: $[-,-,450,103,81,134,406]$  &       &       \\\hline
    SO-LAT  & T: $30-3000$ & T: $[61, 30, 5.3, 6.6, 15, 35]$ & $0.4$ & $[7.4, 5.1, 2.2, 1.4, 1.0, 0.9]$\\
              & P: $30-5000$ & P: $[86.2, 42.4, 7.5, 9.3, 21.2, 49.5]$  &       &       \\\hline
      CMB-S4  & T: $30-3000$ & T: $1.0$ & $0.4$ & $1.5$\\
              & P: $30-5000$ & P: $1.414$ &  &       \\\hline
      CV-limited & T/P: $2-5000$ & $0.0$ & $1.0$ & N/A    \\\hline
    \end{tabular}
    \caption{\color{black}Experimental parameters for Planck, SO-LAT, and a particular configuration of the CMB-S4 experiment used for the Fisher analysis. We also considered a standalone, full-sky, cosmic-variance limited experiment.}
    \label{tab2}
\end{table*}
\begin{figure*}[htbp]
    \centering
    \includegraphics[width=\linewidth]{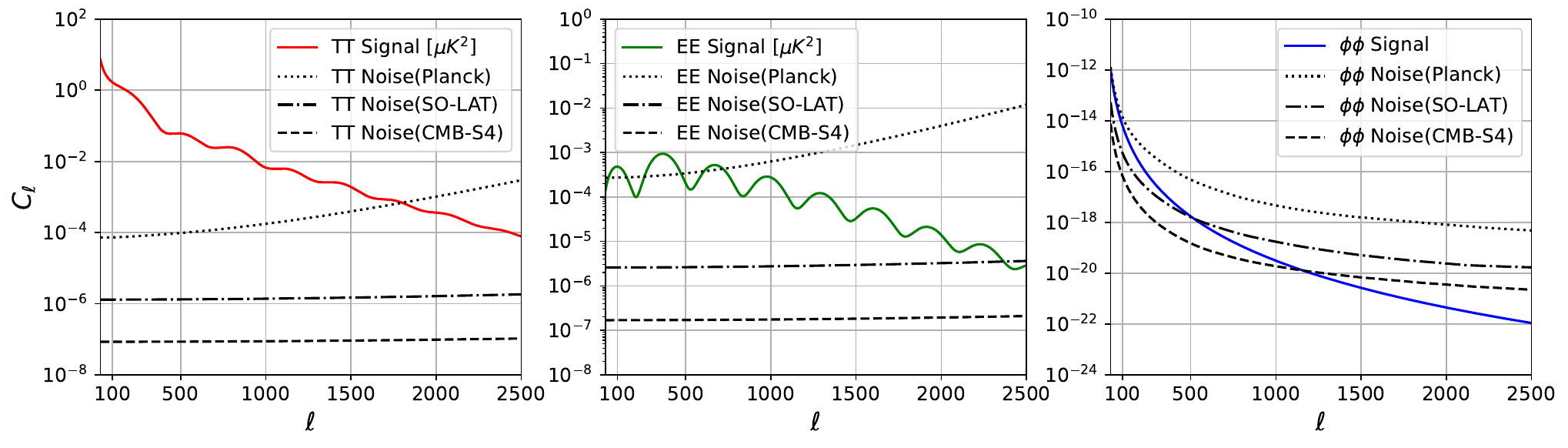}
    \caption{\color{black}Plots of various auto-correlation signal and noise spectra for the inflaton/moduli decay model with the parameter values from Table \ref{tab:tab_fid} (set I).}
    \label{spectra}
\end{figure*}
The $C_\ell$'s appearing in the total covariance matrix $\mathbf C_\ell$ above were obtained from the publicly available package $\mathtt{CLASS}$\footnote{\href{https://github.com/lesgourg/class_public}{https://github.com/lesgourg/class-public}}~\cite{Diego_Blas_2011}, suitably modified to incorporate the non-thermal distribution functions under consideration~\cite{Lesgourgues_2011}. The noise properties for the experiments considered in this work are listed in Table \ref{tab2}~\cite{abazajian2016cmbs4sciencebookedition, Li_2018, SimonsObservatory:2025wwn}, and the TT and EE noise spectra were computed using Eq. (\ref{27}) with these noise parameters. For the lensing convergence, we used the $\mathtt{orphics}$\footnote{\href{https://github.com/msyriac/orphics}{https://github.com/msyriac/orphics}} code~\cite{Madhavacheril_2018} to calculate the noise spectrum $N_\ell^{\kappa\kappa}$ (subsequently converted to $N_\ell^{\phi\phi}$) from a minimum variance combination of $TT,TE,EE,EB,$ and $TB$ quadratic estimators, assuming the CMB B-modes undergo an iterative delensing procedure. Following the CMB-S4 Science Book~\cite{abazajian2016cmbs4sciencebookedition} and the Simons Observatory Science Book (2019)~\cite{Ade_2019}, we have set $\ell_\rm{min}=30$ (recovering large scales from ground is challenging), and $\ell_\rm{max}^T=3000$ and $\ell_\rm{max}^P=5000$ for CMB-S4 and SO-LAT (due to foregrounds). For Planck, we have considered $2<\ell^T<2500$ and $30<\ell^P<2500$. For the lensing signal, we have considered $30<\ell^\phi<2500$. We have assumed infinite noise outside these multipole ranges~\cite{Li_2018} and have set the derivatives of the $C_\ell$'s to zero outside the corresponding multipole ranges to ignore their contribution to our analysis\footnote{See Appendix \ref{appendix:B} for a discussion on numerical derivatives.}. When analyzing $C_\ell^{TT},C_\ell^{TE},C_\ell^{EE}$ in combination with the lensing convergence spectra $C_\ell^{\phi\phi}$ (which has been assumed to contain all the lensing information), we have used unlensed $C_\ell$'s in our Fisher analysis in order to avoid double-counting lensing information~\cite{Li_2018}. 

Measurements of CMB-S4/SO are usually considered in combination with Planck data because the two experiments are highly complementary. Planck, as a satellite mission, provides a higher sky coverage and can also capture the largest angular scales (low $\ell$'s). On the other hand, CMB-S4/SO, being ground-based experiments offers better sensitivity and precision on smaller angular scales (high $\ell$'s). So for the Planck and CMB-S4/SO joint analysis, we added the corresponding Fisher matrices,
\begin{align}
    F=F^\rm{Planck}+F^\rm{CMB-S4/SO},
\end{align}
with the caveat that in the region of overlap between the two experiments, which is $30<\ell<2500$, we set $f_\rm{sky}=0.2$ for Planck in order to avoid double-counting~\cite{Li_2018}. {\color{black}When considering SO in combination with Planck, we set $f_\rm{sky}=0.8$ for Planck for $2<\ell<30$~\cite{Ade_2019}.} We used flat priors for all parameters except $\tau_\rm{reio}$, for which we have added a prior of $0.01$ as follows~\cite{Manzotti_2016}:
\begin{align}
    F_{ii}\to F_{ii}+\frac{1}{\rm{prior}_i^2}.
\end{align}
We utilized the $\mathtt{fishchips}$ package\footnote{\href{https://github.com/xzackli/fishchips-public}{https://github.com/xzackli/fishchips-public}} to develop our code for Fisher analysis.

\section{Results}\label{sec:results}
\begin{figure}[htbp]
    \centering
    \includegraphics[width=\linewidth]{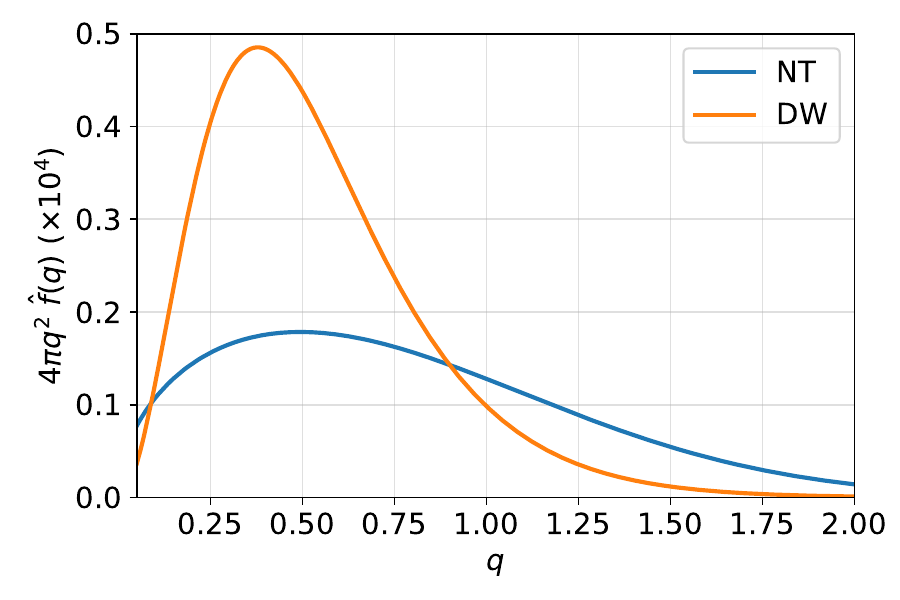}
    \caption{Comparative plot of the inflaton decay (NT) and Dodelson-Widrow (DW) distributions for $\{\Delta N_\rm{eff},M_\rm{sp}^\rm{eff}\}=\{0.034,0.903~\rm{eV}\}$. The momenta and the distribution function are both in units of $T_\rm{ncdm,0}$ which is calculated using Eq. (\ref{Tncdm0}) with $m_\phi=10^{-6}M_\rm{Pl},\tau=10^8/m_\phi,$ and $B_\rm{sp}=0.0118$ for the NT model.}
    \label{fig:compare}
\end{figure}

The fiducial model consists of the following six $\La$CDM parameters: the physical baryon density $\w_b$, the physical cold dark matter density $\w_\rm{cdm}$, the reduced Hubble constant $h$, the amplitude of primordial scalar perturbations $A_s$, the scalar spectral tilt $n_s$, and the optical depth to reionization $\tau_\rm{reio}$, and the two phenomenological parameters $\{\Delta N_\rm{eff}, M_\rm{sp}^\rm{eff}\}$. The phenomenological parameters are however derived parameters --- related to the inflaton/moduli-decay-model parameters $\{B_\rm{sp},m_\rm{sp}\}$ (via Eqs. (\ref{DNeff}) and (\ref{Meff})) or the Dodelson-Widrow-model parameters $\{\chi,m_\rm{sp}\}$ (via Eq. (\ref{DWparams})). For both these models, we assume Standard Model neutrinos to be massless as in Ref.~\cite{Das:2021pof}, since the effect of neutrino masses is expected to be subdominant for the cosmological observables of interest. In Fig. \ref{fig:compare}, we plot the non-thermal (NT) distribution along with the Dodelson-Widrow (DW) distribution for the same values of the phenomenological parameters, viz. $\Delta N_\rm{eff}=0.034$ and $M_\rm{sp}^\rm{eff}=0.903~\rm{eV}$. 
\begin{table}[htbp]
    \centering
    \begin{tabular}{|c|c|c|}
        \hline
        \multirow{2}{*}{Parameter} &  \multicolumn{2}{c|}{Fiducial value}\\
        \cline{2-3}
                  & Set I & Set II\\
                  
        \hline
        $\omega_b$     & 0.02247        & 0.02242  \\
        $\omega_{\rm{cdm}}$ & 0.111   & 0.120 \\
        $h$            & 0.6804          & 0.6711  \\
        $10^9 A_s$     & 2.099           & 2.110 \\
        $n_s$          & 0.9661         &  0.9652  \\
        $\tau_{\rm{reio}}$ & 0.0536   &  0.0560  \\\hline   
        $\Delta N_\rm{eff}$ & 0.034        &   0.1  \\
        $M^\rm{eff}_\rm{sp}$~[eV] & 0.903       &   0.415  \\
        \hline
    \end{tabular}
    \caption{Fiducial values used for Fisher analysis.}
    \label{tab:tab_fid}
\end{table}
\begin{figure}[htbp]
    \centering
    \includegraphics[width=0.9\linewidth]{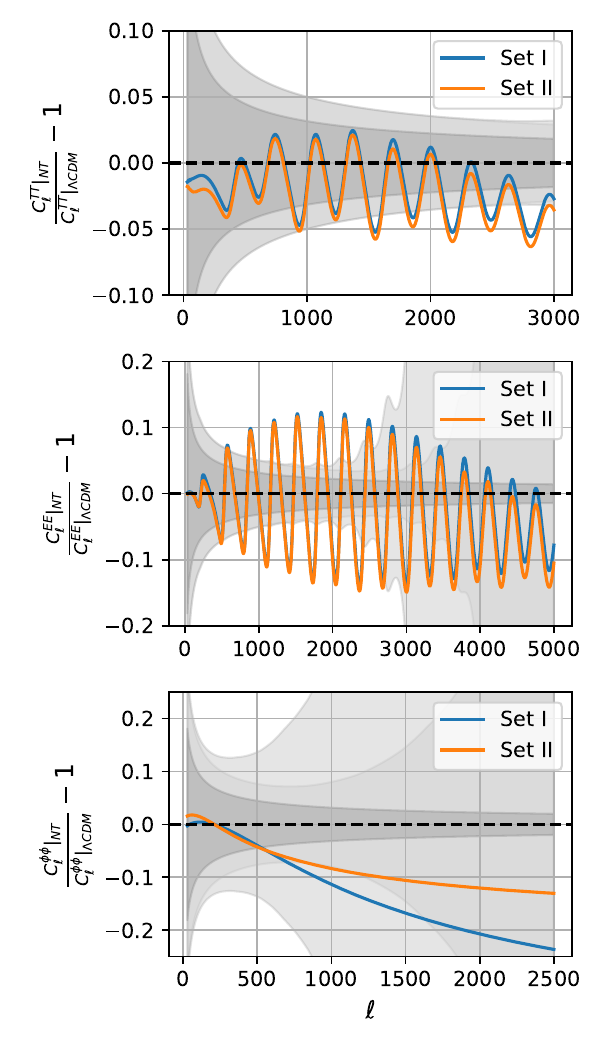}
    \caption{\color{black}Comparison of the residuals of the TT, EE, and $\phi\phi$ spectra for the two sets of fiducial values. The progressively darker shaded regions correspond respectively to the $1\sigma$ error bars associated with the particular configurations of the SO-LAT experiment, the CMB-S4 experiment, and to the CV-limited experiment considered in this work (see Table~\ref{tab2}).}
    \label{fig:Cl_NT}
\end{figure}

We have performed the Fisher analyses with two sets of fiducial values:
\begin{itemize}
    \item Set I: We used the best fit of MCMC analysis with Planck 18+S8 data sets performed in Ref.~\cite{Das:2021pof} for the inflaton decay model. The fiducial values of the $\nu_\rm{NT}\La$CDM model parameters, corresponding to the fiducial values of the phenomenological parameters, $\{\Delta N_\rm{eff},M_\rm{sp}^\rm{eff}\}=\{0.034,0.903~\rm{eV}\}$, are
    \begin{align}
        B_\rm{sp}=0.0118,\quad m_\rm{sp}=38.62~\rm{eV},
    \end{align} 
    whereas for the Dodelson-Widrow model, we have 
    \begin{align}
        \chi=0.034,\quad m_\rm{sp}=26.43~\rm{eV}.
    \end{align}
    \item Set II: We used the mean values of the $\La$CDM parameters obtained from Planck TT, TE, EE+low E+lensing when the base-$\La$CDM model is extended by including $M_\rm{sp}^\rm{eff}$ and $N_\rm{eff}$~\cite{Planck:2018vyg}. In this extension of the base $\La$CDM model, the physical mass of a thermal sterile neutrino is $m^\rm{th}_\rm{sp}=(\Delta N_\rm{eff})^{-3/4}M^\rm{eff}_\rm{sp}$. Assuming the prior $m_\rm{sp}^\rm{th}<10~\rm{eV}$, the Planck-2018 paper found the following $2\sigma$ constraints on $\Delta N_\rm{eff}$ and $M^\rm{eff}_\rm{sp}$ from Planck TT, TE, EE+low E+lensing+BAO:
    \begin{align}
        \Delta N_\rm{eff}<0.246\quad\text{and}\quad M^\rm{eff}_\rm{sp}<0.65~\rm{eV}.
    \end{align}
    We chose $\Delta N_\rm{eff}=0.1$ and {\color{black}$M_\rm{sp}^\rm{eff}=0.415~\rm{eV}$}, which corresponds to the following values of the model parameters:
    \begin{align}
        B_\rm{sp}=0.0332,&\quad m_\rm{sp}=6.20~\rm{eV};\\
    \chi=0.1,&\quad {\color{black}m_\rm{sp}=4.15~\rm{eV}}.
    \end{align}
\end{itemize}
{\color{black} We emphasize that both fiducial sets correspond to the same phenomenological parameterization; they are simply two benchmarks representing different parts of the parameter space---heavier particle with a smaller contribution to $\Delta N_\rm{eff}$ (set I) and lighter particle with a higher contribution to $\Delta N_\rm{eff}$ (set II).} The fiducial values of the $\La$CDM parameters along with the phenomenological parameters $\{\Delta N_\rm{eff},M^\rm{eff}_\rm{sp}\}$ are listed in Table \ref{tab:tab_fid}.

\begin{table*}[htbp]
    \centering
    \begin{tabular}{|c|c|c|c|c|c|c|c|c|}
    \hline
     Parameter    & $10^5\w_b$  & $10^3\w_\rm{cdm}$  & $H_0$  & $\ln{(10^{10}A_s)}$  & $n_s$  & $\tau_\rm{reio}$  & $\Delta N_\rm{eff}$ & $M^\rm{eff}_\rm{sp}~[\rm{eV}]$ \\\hline\hline
     $1\sigma$ error(SO) & $4.20$  & $3.11$  & $0.301$  & $0.011$  & $0.0022$  & $0.0064$  & $0.0061$  & $0.249$  \\\hline
     $1\sigma$ error(Planck+SO) & $3.99$  & $3.05$  & $0.281$  & $0.011$  & $0.0019$  & $0.0061$  & $0.0059$  & $0.247$  \\\hline
     $1\sigma$ error(CMB-S4) & $2.84$  & $1.29$  & $0.270$  & $0.010$  & $0.0020$  & $0.0059$  & $0.0039$  & $0.081$  \\\hline
     $1\sigma$ error(Planck+S4) & $2.74$  & $1.25$  & $0.254$  & $0.010$  & $0.0019$  & $0.0057$  & $0.0038$  & $0.080$  \\\hline
     $1\sigma$ error(CV-limited) & $0.798$  & $0.302$  & $0.092$  & $0.0026$  & $0.0008$  & $0.0014$  & $0.0013$  & $0.0119$  \\\hline
    \end{tabular}
    \caption{\color{black}Projected $1\sigma$ errors for parameters $\theta_i$, marginalized over all other parameters for the inflaton-decay NT model with $\{\Delta N_\rm{eff},M_\rm{sp}^\rm{eff}\}=\{0.034,0.903~\rm{eV}\}$ (set I).}
    \label{tab:fid1sigma}
\end{table*}

\begin{table*}[htbp]
    \centering
    \begin{tabular}{|c|c|c|c|c|c|c|c|c|}
    \hline
     Parameter    & $10^5\w_b$  & $10^3\w_\rm{cdm}$  & $H_0$  & $\ln{(10^{10}A_s)}$  & $n_s$  & $\tau_\rm{reio}$  & $\Delta N_\rm{eff}$ & $M^\rm{eff}_\rm{sp}~[\rm{eV}]$ \\\hline\hline
     $1\sigma$ error(SO) & $5.07$  & $1.42$  & $0.322$  & $0.014$  & $0.0029$  & $0.0072$  & $0.039$  & $0.049$  \\\hline
     $1\sigma$ error(Planck+SO) & $4.95$  & $1.34$  & $0.299$  & $0.013$  & $0.0026$  & $0.0069$  & $0.037$  & $0.047$  \\\hline
     $1\sigma$ error(CMB-S4) & $3.53$  & $1.04$  & $0.285$  & $0.013$  & $0.0024$  & $0.0067$  & $0.030$  & $0.028$  \\\hline
     $1\sigma$ error(Planck+S4) & $3.47$  & $0.99$  & $0.267$  & $0.012$  & $0.0023$  & $0.0065$  & $0.029$  & $0.026$  \\\hline
     $1\sigma$ error(CV-limited) & $0.978$  & $0.277$  & $0.093$  & $0.0028$  & $0.0008$  & $0.0015$  & $0.0066$  & $0.0091$  \\\hline
    \end{tabular}
    \caption{\color{black}Same as Table III with $\{\Delta N_\rm{eff},M_\rm{sp}^\rm{eff}\}=\{0.1,0.415~\rm{eV}\}$ (set II).}
    \label{tab:fid2sigma}
\end{table*}

Various auto-correlation spectra obtained from \texttt{CLASS} (and \texttt{orphics}) using the parameter values of set I for the inflaton/moduli decay model are plotted in Fig. \ref{spectra}, which also shows the improvements in the noise levels for CMB-S4/SO as compared to Planck. In Fig.~\ref{fig:Cl_NT} we show the comparative plots of $C_\ell^{TT}$, $C_\ell^{EE}$, and $C_\ell^{\phi\phi}$ for the two sets of fiducial values for the NT model.  More $\Delta N_\rm{eff}$ leads to more damping of the primary CMB spectra at higher $\ell$. Although the free-streaming velocity of the LiMR is more in the case of set II, the suppression of the lensing spectrum $C_\ell^{\phi\phi}$ is less compared to set I. This is because, in set I, the CDM component is reduced at the cost of the LiMR component. The net effect is less clustering power and hence greater suppression of $C_\ell^{\phi\phi}$. For set II, structure grows more efficiently, despite the fact that $\l<v_\rm{fs}\r>$ is roughly ten times greater than that for set I, leading to stronger lensing potentials and hence $C_\ell^{\phi\phi}$ is less suppressed. We will restrict our analyses to the linear regime since non-linear matter power spectrum corrections (via HALOFIT or HMcode) are not appropriate for models where the dark sector includes massive hot dark matter or warm dark matter components.

Before analyzing these two sets, as a preliminarily study, we will compare
the two non-thermal models, with the LiMR masses fixed to zero, with
the $\La$CDM model extended by the $N_\rm{eff}$ parameter, which we will refer to as the $\n\La$CDM model. We will perform the Fisher analysis by marginalizing over the six $\La$CDM parameters, along with $N_\rm{eff}$ for the $\n\La$CDM model, or $B_\rm{sp}$ for the NT model, or $\chi$ for the DW model, in order to find the {\color{black}projected uncertainties} on $N_\rm{eff}$ around $\Delta N_\rm{eff}=0$. To this end, we use the following fiducial values for the seventh parameter: $N_\rm{eff}=0$, $B_\rm{sp}=0$, and $\chi=0$ for the $\n\La$CDM model, the NT model, and the DW model, respectively\footnote{For the numerical implementation, we set $B_\rm{sp}=10^{-6}$ and $\chi=2.92\times10^{-6}$.}. We use the following fiducial values for the $\La$CDM model parameters~\cite{Planck:2018vyg}: $\omega_b=0.02224,~\omega_\rm{cdm}=0.1179,~h=0.663,~A_s=2.082\times10^{-9},~n_s=0.9589,$ and $\tau_\rm{reio}=0.05$. The $1\sigma$ projected uncertainties on $N_\rm{eff}$ from CMB-S4 were found to be as follows:
\begin{align}\label{Neff0}
    \sigma(N_\rm{eff})=\begin{cases}
    0.028\qquad(\n\La\rm{CDM})\\
    0.025\qquad~~~(\rm{NT})\\
    0.029\qquad~~~(\rm{DW})
    \end{cases}
\end{align}
The fact that {\color{black}our forecasts} for these three models (when expanding around the same fiducial cosmology) {\color{black}are in line with each other} is consistent with the expectation that these models would be indistinguishable since they have the same phenomenological parameters $\Delta N_\rm{eff}\approx0$ and $M_\rm{sp}^\rm{eff}=0~\rm{eV}$.

\begin{table*}[htbp]
    \centering
    \begin{tabular}{|c|c|c|c|c|c|c|c|c|c|}
    \hline
         & Model & $10^5\w_b$  & $10^3\w_\rm{cdm}$  & $H_0$  & $\ln{(10^{10}A_s)}$  & $n_s$  & $\tau_\rm{reio}$  & $\Delta N_\rm{eff}$ & $M^\rm{eff}_\rm{sp}~[\rm{eV}]$ \\\hline\hline
     \multirow{2}{*}{Set I} & NT & $2.85$  & $1.26$  & $0.263$  & $0.010$  & $0.0019$  & $0.0059$  & $0.0038$  & $0.080$  \\\cline{2-10}
     & DW & $2.85$  & $1.17$  & $0.266$  & $0.010$  & $0.0020$  & $0.0059$  & $0.0034$  & $0.073$  \\\hline
     \multirow{2}{*}{Set II} & NT & $3.53$  & $1.03$  & $0.285$  & $0.013$  & $0.0024$  & $0.0067$  & $0.030$  & $0.028$  \\\cline{2-10}
     & DW & $3.73$  & $1.04$  & $0.295$  & $0.013$  & $0.0026$  & $0.0071$  & $0.031$  & $0.030$  \\\hline
    \end{tabular}
    \caption{\color{black}Comparison of the projected $1\sigma$ errors from CMB-S4 for parameters $\theta_i$, marginalized over all other parameters for the NT and DW models.}
    \label{tab:fidsigmaNTDW}
    \end{table*}

\subsection{Forecasts for inflaton-decay non-thermal and Dodelson-Widrow models}
Now we move on to the results of the Fisher forecasts for the LiMRs with the fiducial values in Table \ref{tab:tab_fid}.
\begin{itemize}
    \item Set I: The projected $1\sigma$ uncertainties on   the fiducial model parameters are reported in Table \ref{tab:fid1sigma}. The Fisher ellipses for the parameters $\{\w_m=\w_b+\omega_\rm{cdm},h,\Delta N_\rm{eff},M_\rm{sp}^\rm{eff}\}$ are shown in Fig. \ref{fig:nt2small}.
    As expected, combined analysis of Planck and CMB-S4/SO leads to tighter constraints on the parameters. From Table \ref{tab:fidsigmaNTDW}, one can see that the NT model and the DW model uncertainties are in close agreement with each other.
    Of our particular interest are the phenomenological parameters, whose $1\sigma$ uncertainties from CMB-S4+Planck are
    \begin{align}\label{Neff2}
        1\sigma(\Delta N_\rm{eff})=\begin{cases}
        0.0038\quad(\rm{NT})\\
        0.0033\quad(\rm{DW})
        \end{cases},
    \end{align}
    and
    \begin{align}\label{Meff2}
        1\sigma(M_\rm{sp}^\rm{eff})=\begin{cases}
        80~\rm{meV}\quad(\rm{NT})\\
        76~\rm{meV}\quad(\rm{DW})
        \end{cases}.
    \end{align} 
    
    \item Set II: The Fisher plots are shown in Fig. \ref{fig:nt4small} and the $1\sigma$ uncertainties are given in Table \ref{tab:fid2sigma}. We again see from Table \ref{tab:fidsigmaNTDW} that the DW model parameter uncertainties agree quite well with the corresponding uncertainties for the NT model. {\color{black}With this set of fiducial values, we obtain the following $1\sigma$ uncertainties from CMB-S4+Planck:}
    \begin{align}\label{Neff4}
        1\sigma(\Delta N_\rm{eff})=\begin{cases}
        0.029\quad(\rm{NT})\\
        0.030\quad(\rm{DW})
        \end{cases},
    \end{align}
    and 
    \begin{align}\label{Meff4}
        1\sigma(M_\rm{sp}^\rm{eff})=\begin{cases}
        26~\rm{meV}\quad(\rm{NT})\\
        29~\rm{meV}\quad(\rm{DW})
        \end{cases}.
    \end{align}
\end{itemize} 

\begin{figure*}[htbp]
    \centering
    \begin{subfigure}[t]{0.48\textwidth}
        \centering
        \includegraphics[width=\textwidth]{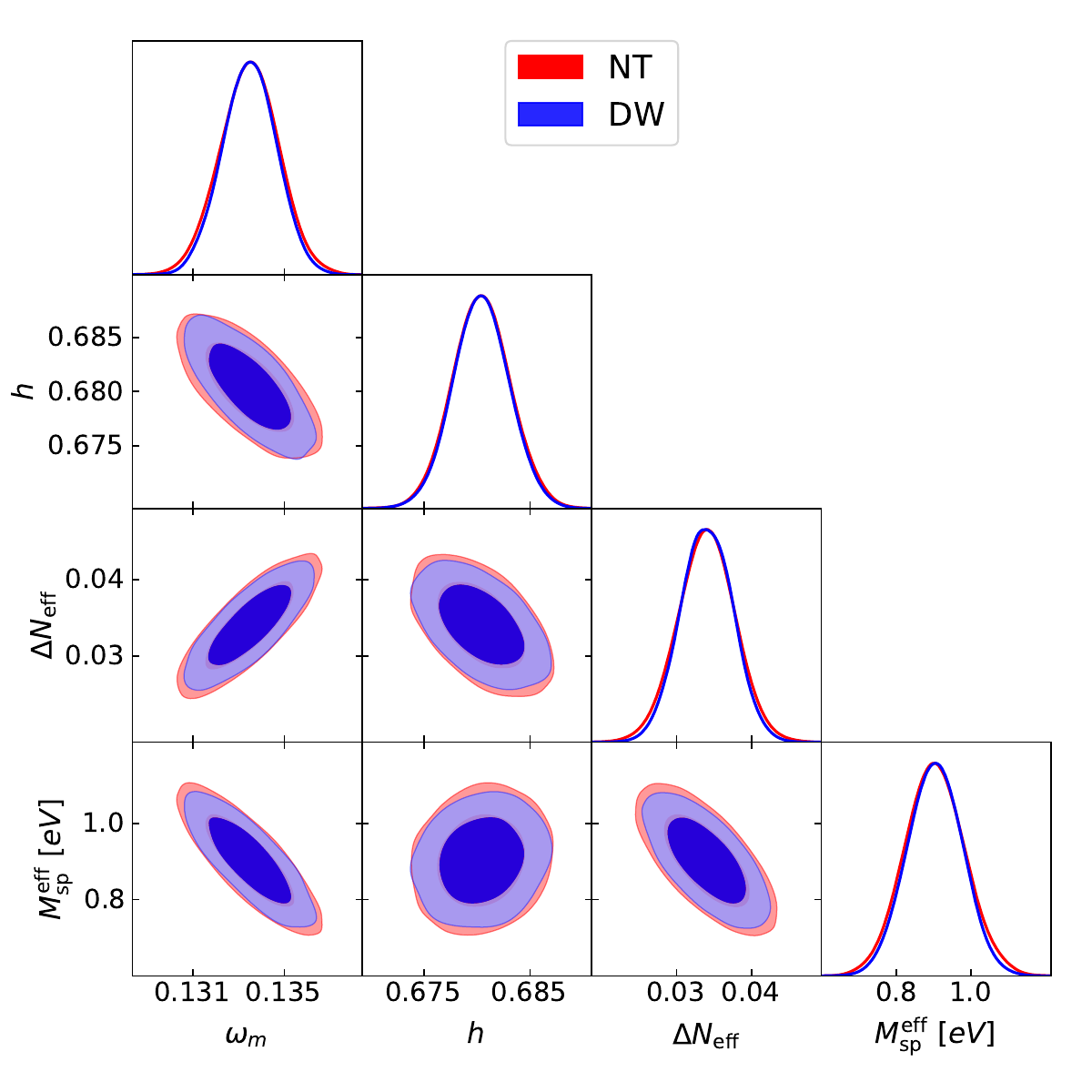}
        \caption{Fisher plots for the NT and DW models with fiducial values set I.}
    \label{fig:nt2small}
    \end{subfigure}
    \hfill
    \begin{subfigure}[t]{0.48\textwidth}
        \centering
        \includegraphics[width=\textwidth]{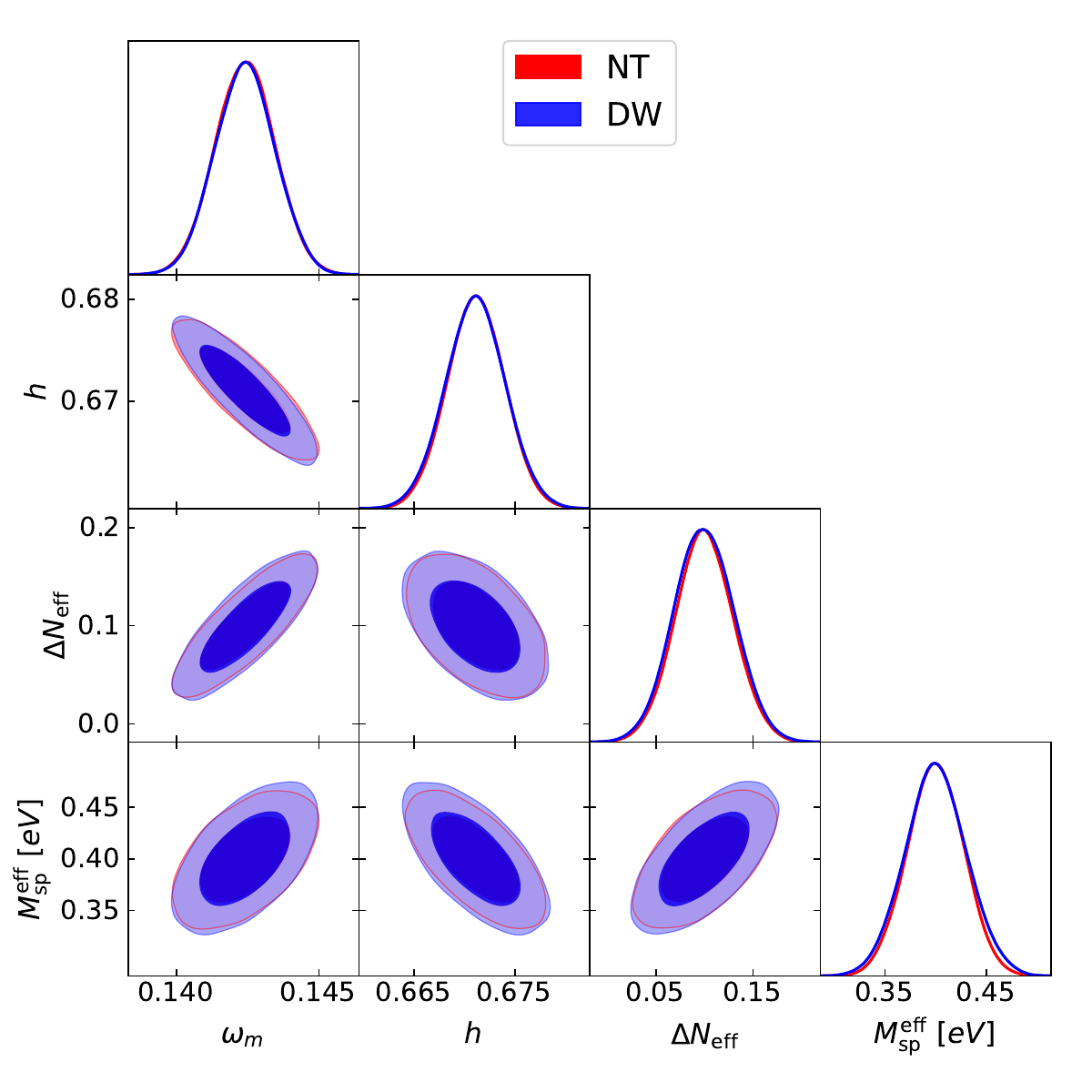}
        \caption{Same as (a) with fiducial values set II.}
    \label{fig:nt4small}
    \end{subfigure}
    \caption{\color{black}The posterior distributions and Fisher ellipses for the parameters $\{\omega_m=\w_b+\w_\rm{cdm},h,\Delta N_\rm{eff},M_\rm{sp}^\rm{eff}\}$ (plotted using \texttt{getdist}) for the CMB-S4 experiment.}
    \label{fig:fisher_ellipses}
\end{figure*}
\begin{figure}[htbp]
    \centering
    \includegraphics[width=\linewidth]{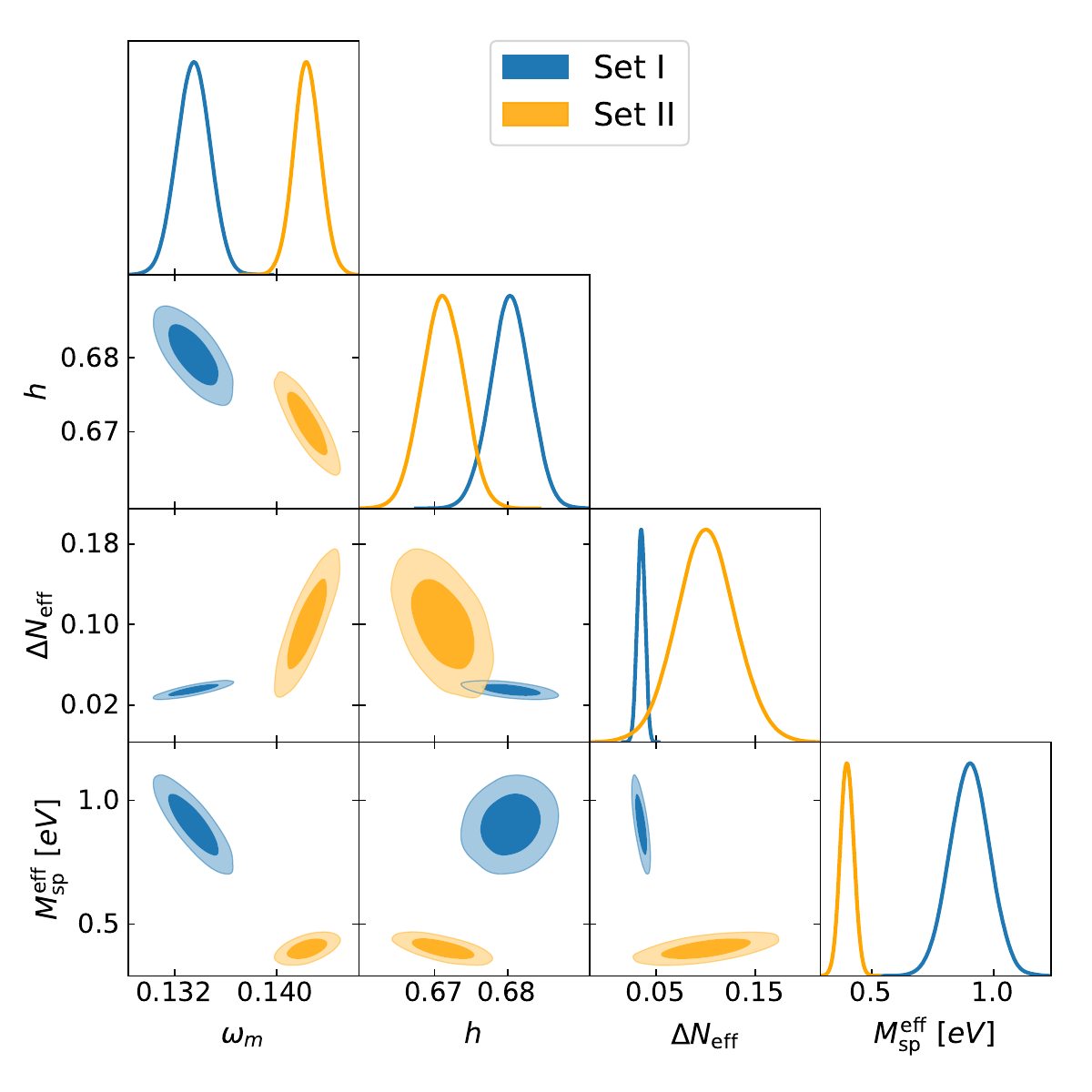}
    \caption{Comparative Fisher plots for CMB-S4.}
    \label{fig:fisherS4}
\end{figure}
\begin{figure}[htbp]
    \centering
    \includegraphics[width=\linewidth]{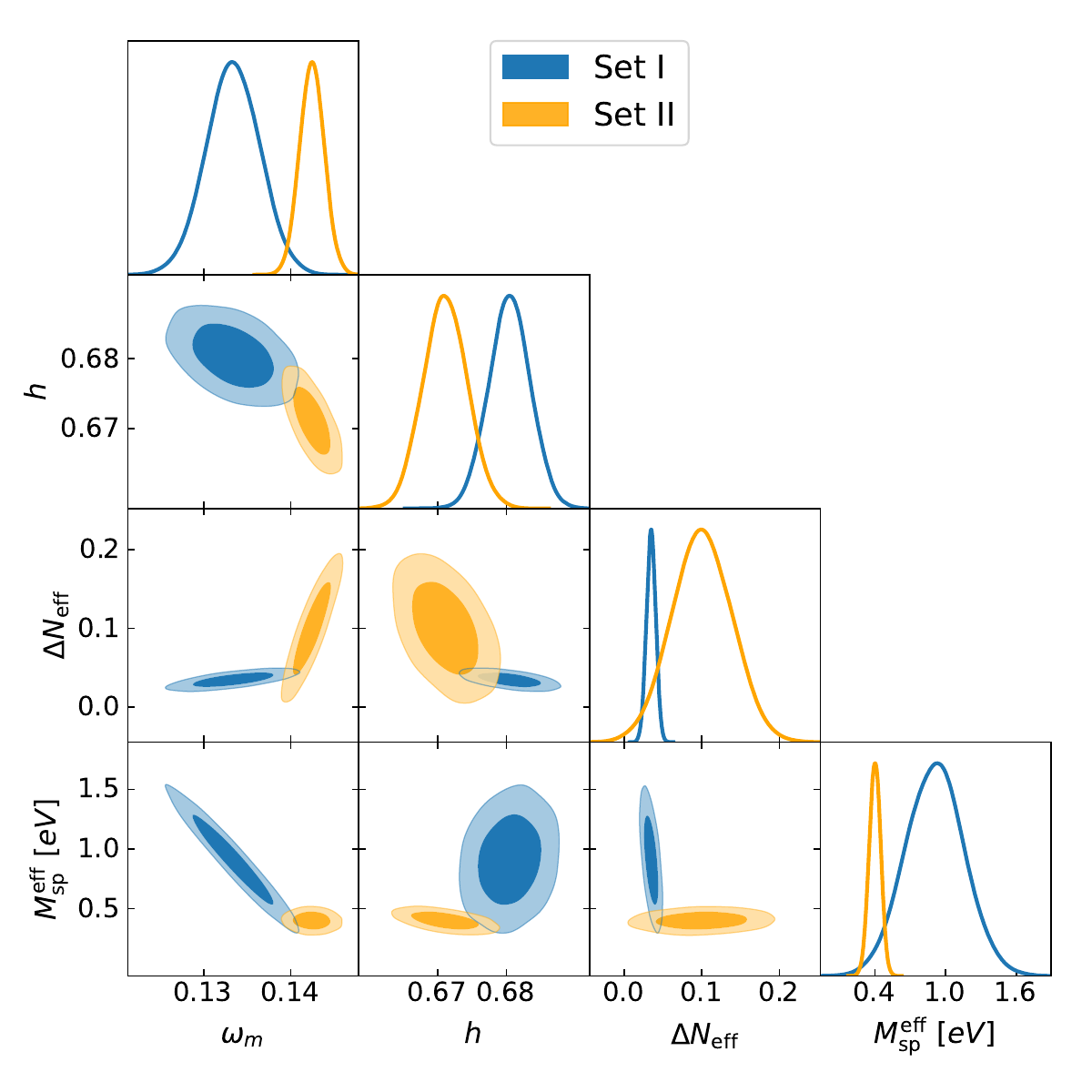}
    \caption{\color{black}Comparative Fisher plots for SO-LAT.}
    \label{fig:fisherSO}
\end{figure}
{\color{black} We see that the relative uncertainties on $M_\rm{sp}^\rm{eff}$ are $8.9\%(8.3\%)$ and $6.3\%(7\%)$, whereas the relative uncertainties on $\Delta N_\rm{eff}$ are $11.2\%(9.7\%)$ and $29\%(30\%)$, for the NT(DW) model. So the relative uncertainties on $M_\rm{sp}^\rm{eff}$ slightly worsen while there is a significant improvement in the relative uncertainties on $\Delta N_\rm{eff}$ in going from set II to set I. Note that the Fisher analysis has been implemented via the fundamental parameterization---$\{B_\rm{sp},m_\rm{sp}\}$ for the NT model and $\{\chi,m_\rm{sp}\}$ for the DW model---and then projected uncertainties have been translated to those on the phenomenological parameters (using Eq. (\ref{jac})). In Figs.~\ref{fig:fisherS4}, and~\ref{fig:fisherSO} we compare the Fisher ellipses and the posterior distributions obtained from the CMB-S4, and SO-LAT for the two sets of parameters considered in the text.}

\subsection{Discussion}\label{sec:discussion}

Our results reveal the following  noteworthy features pertaining to the phenomenological parameters.
\begin{itemize}  
 \item[(i)] There is significant difference in the relative uncertainties  on $\Delta N_{\rm{eff}}$ for the two sets analyzed (as seen in Eq.~(\ref{Neff2}) and Eq.~(\ref{Neff4})).
 \item[(ii)]  The phenomenological parameters, $\Delta N_\rm{eff}$ and $M_\rm{sp}^\rm{eff}$, are negatively correlated for the heavier particle but positively correlated for the lighter one.
\end{itemize}
In this section, we will discuss these features and provide explanations for them. 

\begin{figure}[htbp]
    \centering
        \begin{subfigure}{0.48\textwidth}
        \centering
        \includegraphics[width=\textwidth]{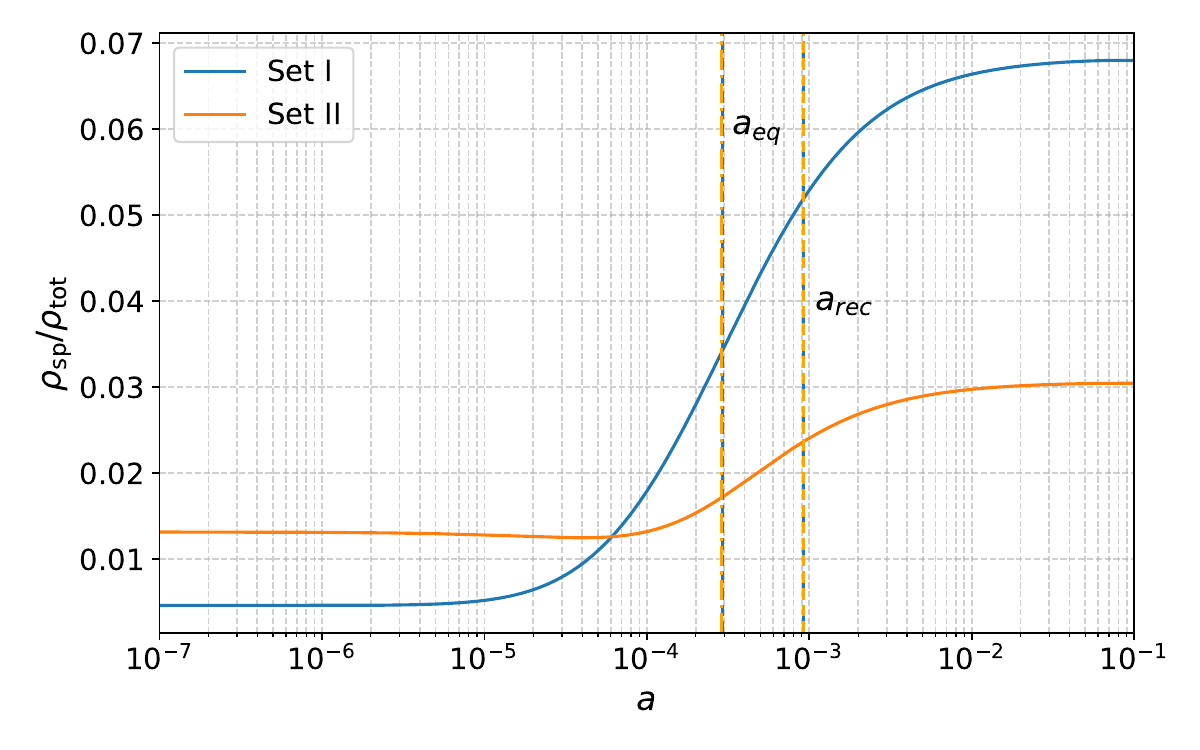}
        \caption{Evolution of the energy density $\rho_\rm{sp}$ of the particle, as a fraction of the total energy density $\rho_\rm{tot}$ of the Universe, with the scale factor.}
    \label{fig:rho}
    \end{subfigure}
    \label{fig:eos_rho}
    \hfill
    \begin{subfigure}{0.48\textwidth}
        \centering
        \includegraphics[width=\textwidth]{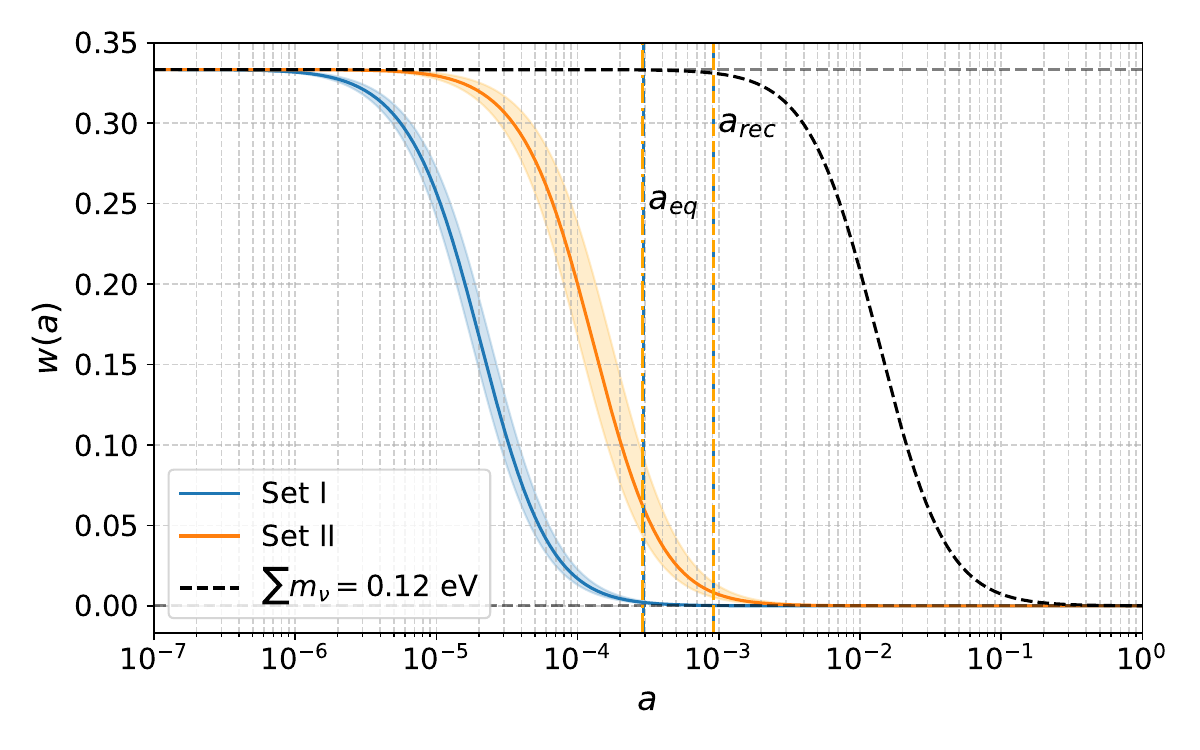}
        \caption{$w$ vs $a$ plot for the LiMRs corresponding to the two sets of fiducial values. For reference, we have also shown the behavior of massive standard (degenerate) neutrinos with $\sum m_\n=0.12~\rm{eV}$. The shaded regions correspond to the $1\sigma$ uncertainties on $w$ from CMB-S4.}
    \label{fig:eos}
    \end{subfigure}
    \caption{Comparison of the evolution of the equation of state parameter and the energy density of the LiMR for the two sets of fiducial values (for the NT model).}
\end{figure}

\begin{figure*}[htbp]
\centering

\centering
\begin{tikzpicture}[scale=0.85,
    box/.style={
        rectangle, draw, rounded corners,
        align=center, minimum width=6cm,
        inner sep=4pt, font=\small
    },
    arrow/.style={->, thick},
    node distance=0.8cm
]

\node[font=\large\bfseries] (titleI) at (0,0) {Set I};
\node[font=\small] (subtitleI) [below=1mm of titleI] 
    {(LiMR fully non-relativistic at $a_\rm{eq}$)};
    
\node[box] (a1) at (0,-1.8) {Increase $\Delta N_\rm{ eff}$ (via $B_\rm{sp}$ or $\chi$)\\\footnotesize{(Adds LiMR matter at equality + recombination)}};
\node[box, below=of a1] (b1) {$\w_\rm{sp}\uparrow,~\w_r\sim\implies z_\rm{eq} \uparrow$\\$D_{A*}\downarrow$ more than $r_{s*}\downarrow\implies\theta_*\uparrow$};
\node[box, below=of b1] (c1) {$M_\rm{sp}^\rm{eff}$ decreases via $m_\rm{sp}$\\\footnotesize{(Removes excess LiMR)}\\$z_\rm{eq}\downarrow$ and $\theta_*\downarrow$};
\node[box, below=of c1] (d1) {$\w_m$ increases\\$z_\rm{eq}\uparrow$ (restored) and $\theta_*\downarrow$};
\node[box, below=of d1] (e1) {$h$ decreases\\ $z_\rm{eq}\sim$ and $\theta_*\uparrow$ (restored)};

\node[box, below=10mm of e1, minimum width=4cm] (result) {
$\Delta N_\rm{eff}$ correlations with\\
$M_\rm{sp}^\rm{eff}$: negative\\
$\omega_m$: positive\\
$h$: negative
};

\draw[arrow] (a1) -- (b1);
;
\draw[arrow] (b1) -- 
node[midway, right, font=\scriptsize]{primary CMB} (c1);
\draw[arrow] (c1) -- 
node[midway, right, font=\scriptsize]{primary CMB} (d1);
\draw[arrow] (d1) -- 
node[midway, right, font=\scriptsize]{primary CMB+lensing} (e1);

\end{tikzpicture}
\hfill
\centering
\begin{tikzpicture}[scale=0.85,
    box/.style={
        rectangle, draw, rounded corners,
        align=center, minimum width=6cm,
        inner sep=4pt, font=\small
    },
    arrow/.style={->, thick},
    node distance=0.8cm
]

\node[font=\large\bfseries] (titleII) at (0,0) {Set II};
\node[font=\small] (subtitleII) [below=1mm of titleII] 
    {(LiMR partially relativistic at $a_\rm{eq}$)};

\node[box] (a2) at (0,-1.8) {Increase $\Delta N_\rm{eff}$ (via $B_\rm{sp}$ or $\chi$)\\\footnotesize(More radiation-like energy at eq. and rec.)};
\node[box, below=of a2] (b2) {$\w_\rm{sp}\uparrow$ more than $\w_r\uparrow\implies z_\rm{eq}\uparrow$\\$D_{A*}\downarrow$ more than $r_{s*}\downarrow\implies\theta_*\uparrow$};
\node[box, below=of b2] (c2) {$M_\rm{sp}^\rm{eff}$ decreases$^*$ via $m_\rm{sp}$\\\footnotesize{($^*$still greater than its fiducial value)}\\
$z_\rm{eq}\downarrow$ and $\theta_*\downarrow$};
\node[box, below=of c2] (d2) {$\w_m$ increases\\\footnotesize{($\because$ LiMR free-streaming suppresses lensing power)}\\$z_\rm{eq}\downarrow$ (restored) and $\theta_*\downarrow$};
\node[box, below=of d2] (e2) {$h$ decreases\\ $z_\rm{eq}\sim$ and $\theta_*\uparrow$ (restored)};

\node[box, below=5mm of e2, minimum width=4cm] (result) {
$\Delta N_\rm{eff}$ correlations with\\
$M_\rm{sp}^\rm{eff}$: positive\\
$\omega_m$: positive\\
$h$: negative
};

\draw[arrow] (a2) -- (b2);
\draw[arrow] (b2) --
node[midway, right, font=\scriptsize]{primary CMB} (c2);
\draw[arrow] (c2) -- 
node[midway, right, font=\scriptsize]{primary CMB+lensing} (d2);
\draw[arrow] (d2) -- 
node[midway, right, font=\scriptsize]{primary CMB+lensing} (e2);

\end{tikzpicture}
\caption{\color{black}Schematic flowcharts summarizing the physical origin of parameter correlations for Set I (left) and Set II (right) based (primarily) on ($z_\rm{eq},\theta_*$). Labels on the arrows indicate the CMB primary and/or lensing contributions that govern the corresponding step. See Appendix~\ref{appendix:C} for details.
}
\label{fig:summary}
\end{figure*}

First, let us look at bullet point (i).
We will argue that the large difference in {\color{black}the forecasts on} $\Delta N_\rm{eff}$ when considering the two sets of fiducial parameter values within the same model arises from the LiMR's contribution to the energy budget just prior to recombination, in particular their relative contributions to the ``matter" (with energy density scaling as $a^{-3}$) and ``radiation" (energy density scaling as $a^{-4}$) components. The heavier particle (set I), owing to its lower number density, has a relatively lesser contribution to the relativistic energy density, compared to the lighter particle (set II), but it contributes significantly more to the matter density, as demonstrated in Fig. \ref{fig:rho}.  Evolution of the effective equation of state parameter of the LiMRs are plotted in Fig.~\ref{fig:eos}. The heavier particle becomes non-relativistic just before matter-radiation equality. By recombination, it therefore contributes solely to the total matter density. Contrast this with the lighter particle of set II (contributing more to $N_\rm{eff}$), which is caught in transition at matter-radiation equality and becomes completely non-relativistic only after recombination. As such, the CMB becomes sensitive to the heavier relic almost exclusively through its effect on the weak lensing signal. When the lensing signal was excluded from the Fisher computations, we obtained $\sigma(\Delta N_\rm{eff})\sim0.01~(0.04)$ for the heavier (lighter) particle. Even in the absence of lensing, the {\color{black}smaller uncertainty} for set I is due to the fact that the particle contributes solely as extra matter by recombination. Any shift in the total matter density at that epoch alters the acoustic peak structure in a way that $\Delta N_\rm{eff}$ (being zero at recombination) cannot mimic. Since $\Delta N_\rm{eff}$ has no freedom to absorb these matter-like changes, the primary spectra pin down its fiducial value to a higher precision. The effect of lensing information is more pronounced in the case of set I (as mentioned above) which leads to a greater improvement in the uncertainty on $\Delta N_\rm{eff}$ (by a factor of $\sim3$) as compared to set II (where there is a $\sim30\%$ improvement). The heavier LiMR is more degenerate with $\w_m$ at recombination compared to the lighter one -- for a given total matter content (baryons+CDM+LiMR), replacing a fraction of $\w_m$ with $\w_\rm{sp}$ has negligible impact for the heavier particle -- this explains the slightly degraded relative uncertainty on $M_\rm{sp}^\rm{eff}$ for set I as compared to set II. 

{\color{black}Now we move on to bullet point (ii) -- the correlation between the phenomenological parameters. We have summarized the physical origins of the correlations of $\Delta N_\rm{eff}$ with $M_\rm{sp}^\rm{eff}$, as well as with $\w_m$ and $h$ in a schematic flowchart (Fig.~\ref{fig:summary}). The detailed explanations have been relegated to Appendix~\ref{appendix:C}, where we also quantitatively validate our qualitative arguments.}

\begin{figure}[htbp]
    \centering
    \includegraphics[width=0.95\linewidth]{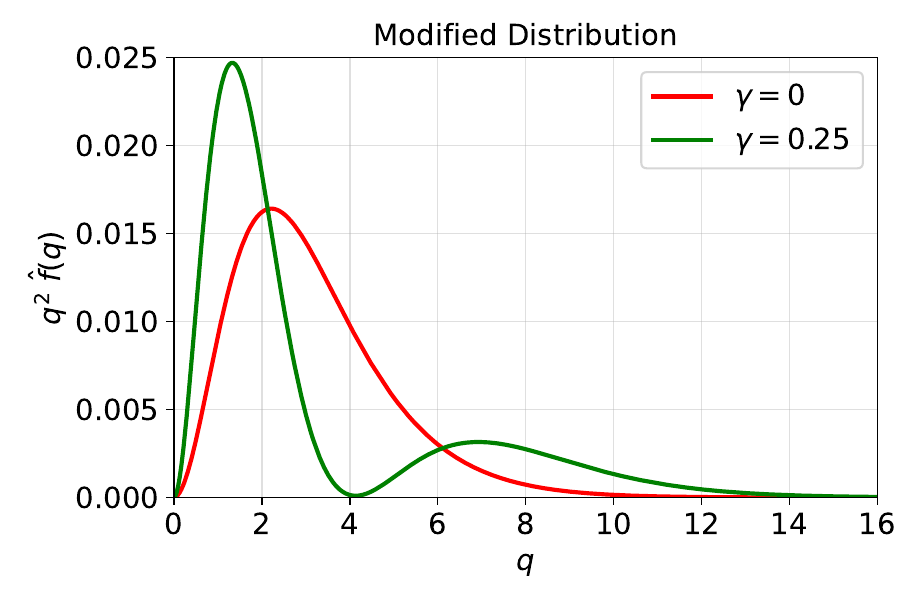}
    \caption{Plot showing the modified distribution function for $\c=0$ (same as DW) and $\c=0.25$, with the momentum expressed in terms of the neutrino temperature today.}
    \label{fig:momentscompare}
\end{figure}
\section{Higher Moments}\label{sec:higher}

So far we discussed the impact of LiMRs in terms of two phenomenological parameters $M_\rm{sp}^\rm{eff}$ and $\Delta N_\rm{eff}$ which are proportional to the zeroth and first moments of the phase space distribution function respectively.  We obtained similar results for both the inflaton-decay model and the Dodelson-Widrow model once $M_\rm{sp}^\rm{eff}$ and $\Delta N_\rm{eff}$ are matched, in spite of the noticeably different forms of the distribution functions (see Fig. \ref{fig:compare}). Ref.~\cite{Acero_2009} claimed that different models with the same $M_\rm{sp}^\rm{eff}$ and $\Delta N_\rm{eff}$ would be difficult to distinguish, especially using observables at the linear perturbation theory level. To check how well this claim holds up at the noise levels expected at the CMB-S4 experiment, we consider two distribution functions with matched zeroth and first moments --- the Dodelson-Widrow distribution $\hat f_{DW}(q)$ and another distribution $\hat f(q)$ which starts differing from the Dodelson-Widrow distribution at the level of the second moment. In order to include the effect of the second moment we consider an Edgeworth-like expansion about the Dodelson-Widrow distribution, which constitutes a continuous deformation of the Dodelson-Widrow distribution:
\begin{align}\label{f1}
    \hat f(q) = \hat f_{DW}(q)\l[\a_0P_0(q)+\a_1P_1(q)+\a_2P_2(q)\r],
\end{align}
where $\hat f_{DW}(q)=\chi[\exp{(q)}+1]^{-1}$. The functions $P_m(q)$ are $m$-th order polynomials in $q$, orthonormal with respect to the measure $q^2/(e^q+1)$~\cite{Cuoco_2005},
\begin{align}
    \int_0^\infty dq\frac{q^2}{e^q+1}P_m(q)P_n(q)=\d_{mn}.
\end{align}
The modified distribution function in (\ref{f1}) can be rewritten as
\begin{align}\label{fq}
    \hat f(q)=\frac{\chi}{e^q+1}\l[\a+\b q+\c q^2\r],
\end{align}
where $\a,\b,$ and $\c$ are dimensionless constants, and its $n$-th moment computed as
\begin{align}
    Q^{(n)}=\int_0^\infty dq~q^{2+n}\hat f(q).
\end{align}
Setting $Q^{(0)}=Q^{(0)}_{DW}$ and $Q^{(1)}=Q^{(1)}_{DW}$, \ie matching the first two moments, we can solve for $\a$ and $\b$ in terms of $\c$ to obtain
\begin{align}\label{finalf}
    \hat f(q)=\frac{\chi}{e^q+1}\l[1 + 13.0551\c - 8.24864\c q + \c q^2\r],
\end{align}
where $\c$ is now a free parameter which determines how much $\hat f(q)$ deviates from $\hat f_{DW}(q)$. However, to be a valid distribution function, $\hat f(q)$ has to be non-negative everywhere. This requirement limits the value of $\c$ to the range $0\leq\c\leq0.25$, where $\c=0$ corresponds to the DW distribution (see Fig. \ref{fig:momentscompare}). That $\hat f(q)$ goes to zero at large $q$ values is ensured by the $1/(e^q+1)$ factor which goes like $e^{-q}$ for large $q$. 

\begin{figure}[htbp]
    \centering
    \begin{subfigure}{0.4\textwidth}
        \centering
        \includegraphics[width=\textwidth]{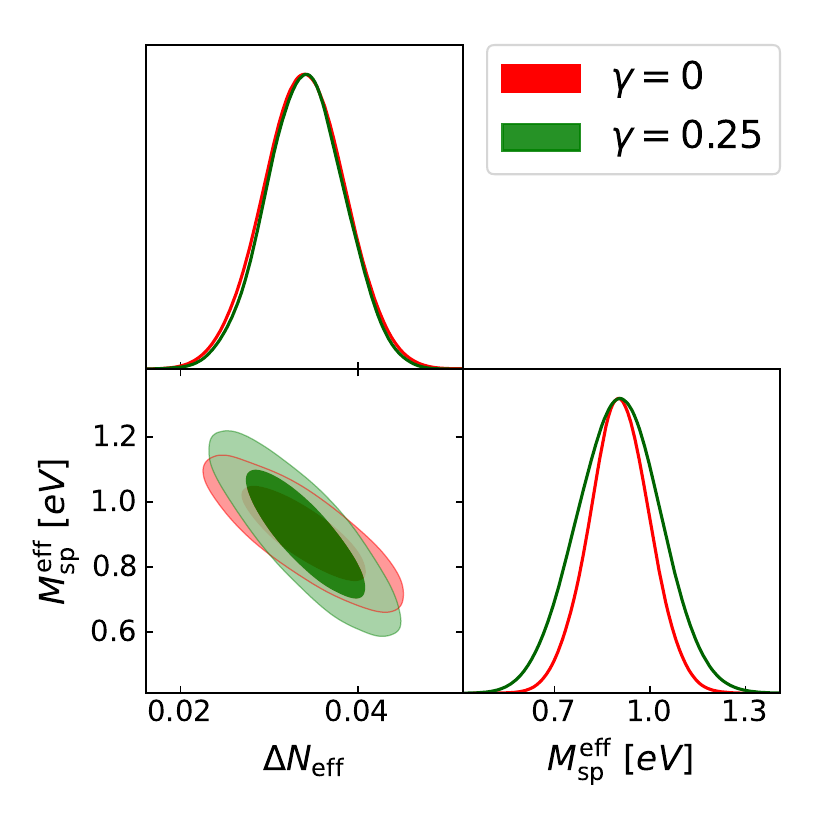}
        \caption{Planck+CMB-S4 forecast.}
    \end{subfigure}
    \hfill
    \begin{subfigure}{0.4\textwidth}
        \centering
        \includegraphics[width=\textwidth]{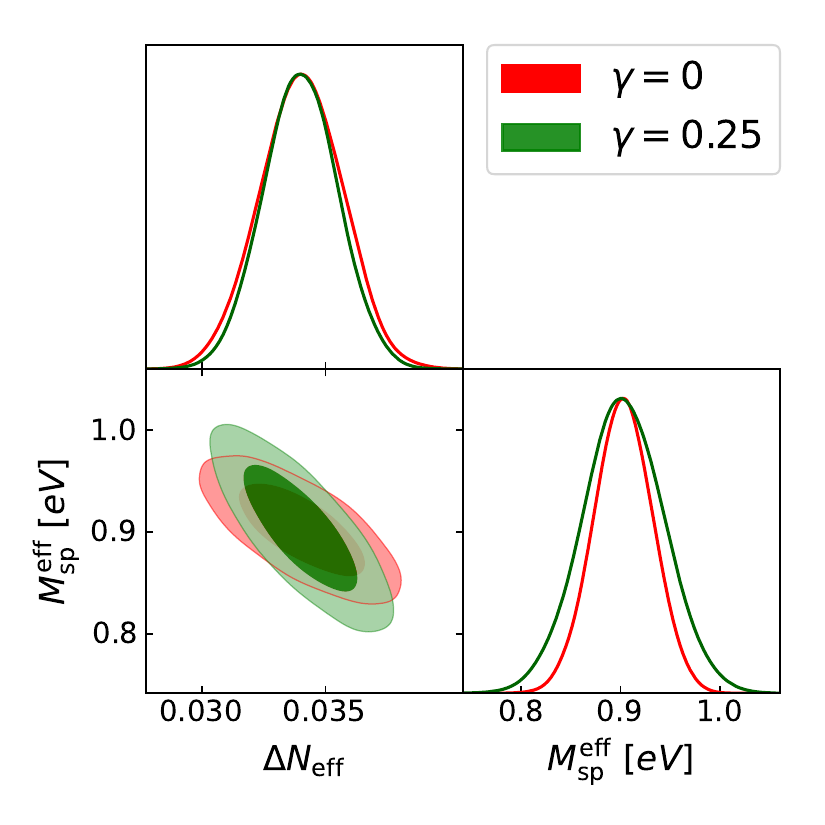}
        \caption{CV-limited experiment forecast.}
    \end{subfigure}
    \caption{{\color{black}Fisher plots of the two phenomenological parameters for the modified distribution.}}
    \label{fig:moments1}
\end{figure}

The projected $1\sigma$ uncertainties on {\color{black}the six $\La$CDM and the two phenomenological parameters}, obtained from our Fisher analysis with the fiducial values set I, for the cases $\c=0$ and $\c=0.25$ are given in Table~\ref{tab:moments1}. The corresponding Fisher ellipses and posterior distributions for the phenomenological parameters are shown in Fig.~\ref{fig:moments1}.
\begin{table*}[htbp]
    \centering
    \begin{tabular}{|c|c|c|c|c|c|c|c|c|c|}
    \hline \multirow{1}{*}{Experiment}
    & Parameter $\rightarrow$   & $10^5\w_b$  & $10^3\w_\rm{cdm}$  & $H_0$  & $\ln{(10^{10}A_s)}$  & $n_s$  & $\tau_\rm{reio}$  & $\Delta N_\rm{eff}$ & $M^\rm{eff}_\rm{sp}~[\rm{eV}]$   \\\hline
    \multirow{2}{*}{Planck+CMB-S4} & $1\sigma$ error $(\c=0)$ &  $2.74$  & $1.14$  & $0.25$  & $0.0099$  & $0.0019$  & $0.0056$  & $0.0033$  & $0.073$  \\\cline{2-10}
    & $1\sigma$ error $(\c=0.25)$ & $2.76$  & $1.41$  & $0.23$  & $0.0099$  & $0.0018$  & $0.0056$  & $0.0036$  & $0.110$ \\\hline
    \multirow{2}{*}{CV-limited} & $1\sigma$ error $(\c=0)$ & $0.802$ & $0.291$ & $0.093$ & $0.0026$ & $0.00083$ & $0.0014$ & $0.0013$ & $0.0121$  \\\cline{2-10}
    & $1\sigma$ error $(\c=0.25)$ & $0.806$ & $0.292$ & $0.085$ & $0.0026$ & $0.00083$ & $0.0014$ & $0.0011$ & $0.0133$ \\\hline
    \end{tabular}
    \caption{\color{black}Projected uncertainties on the parameters for the modified DW distribution with fiducial values set I.}
    \label{tab:moments1}
\end{table*}
\begin{table}[htbp]
    \centering
    \begin{tabular}{|c|c|c|c|}
    \hline
    Experiment  & $\Delta N_\rm{eff}$ & $M_\rm{sp}^\rm{eff}~\rm{[eV]}$ & $\c$ \\\hline
    \multirow{2}{*}{CMB-S4} & $0.034\pm0.0047$ & $0.903\pm0.097$ & $0.00^{+0.175}$  \\\cline{2-4}
    & $0.034\pm0.0044$ & $0.903\pm0.129$ & $0.25_{-0.168}$ \\\hline
    \multirow{2}{*}{CV-limited} & $0.034\pm0.0014$ & $0.903\pm0.022$  & $0.00^{+0.034}$ \\\cline{2-4}
    & $0.034\pm0.0013$ & $0.903\pm0.033$ & $0.25_{-0.063}$
    \\\hline
    \end{tabular}
    \caption{\color{black}Projected uncertainties on the parameters $\{\Delta N_\rm{eff},M_\rm{sp}^\rm{eff},\c\}$ for the modified DW distribution. Note that the uncertainties on $\c$ are one-sided.}
    \label{tab:compgamma}
\end{table}
To determine whether the CMB-S4 experiment will be able to distinguish between these two models, we performed Fisher forecasts by marginalizing over the six $\La$CDM parameters, the two phenomenological parameters $\Delta N_\rm{eff}$ and $M_\rm{sp}^\rm{eff}$, along with the $\gamma$ parameter. In Table~\ref{tab:compgamma} we show the fiducial values and the projected uncertainties for $\Delta N_\rm{eff},M_\rm{sp}^\rm{eff},$ and $\c$. {\color{black}Note that increasing the number of parameters being varied worsens the projected uncertainty on each parameter (compare with Table~\ref{tab:moments1}).}
We find that for $\c=0$, the modified distribution with $\c=0.25$ is $\sim1.4\sigma$ away, and for $\c=0.25$, the DW distribution ($\c=0$) is $\sim1.5\sigma$ away, when we consider the posterior probability distribution of the $\c$ parameter. This indicates that it is unlikely that CMB-S4 would be able to distinguish between the two distributions at a statistical significance. We believe this generalizes to any two LiMR models having the same values of $\Delta N_\rm{eff}$ and $M_\rm{sp}^\rm{eff}$ irrespective of how much the higher moments differ. However, for the CV limited experiment, the two models were found to be distinguishable (see Table~\ref{tab:compgamma}):
when assuming the $\c=0$ case to be the true model, we found the model corresponding to the $\c=0.25$ case to lie outside the $\sim7.4\sigma$ C.L., whereas when considering the latter model to be the true one, the former was found to lie outside the $\sim3.9\sigma$ C.L. So a future experiment with lower noise levels (\ie better sensitivity) than the CMB-S4 experiment, for instance the CMB-HD experiment~\cite{sehgal2020cmbhdastro2020rfiresponse}, can potentially differentiate between two models with the same first two moments. 
Note that the uncertainties on $\c$ have been obtained using Eq. (\ref{sigma_i}), and we have not implemented the hard prior on $\c$, \ie $0\leq\c\leq0.25$. 

In order to incorporate the valid range of $\c$ in our Fisher analysis, we computed the covariance matrix by numerically sampling a multivariate Gaussian from the inverse of the Fisher matrix and rejecting those realizations for which $\gamma$ lies outside the allowed range. {\color{black}We obtained the following median values and asymmetric uncertainties (due to the non-Gaussian nature of the posterior distribution induced by the hard prior; see footnote 2):}
\begin{align}
    \c=0.098^{+0.088}_{-0.068}\quad\text{and}\quad\c=0.154^{+0.066}_{-0.088},
\end{align}
corresponding to the fiducial values $\c=0$ and $\c=0.25$ respectively. Even with the imposition of the hard priors, which leads to {\color{black}more precise (asymmetric) forecasts}, the distribution with $\c=0.25$ is within the $3\sigma$ uncertainty threshold of the $\c=0$ distribution and vice versa. This shows that including the hard prior does not change our conclusion above and moving on we shall use the one-sided uncertainties on $\c$ obtained previously.
\begin{figure}
    \centering
    \includegraphics[width=\linewidth]{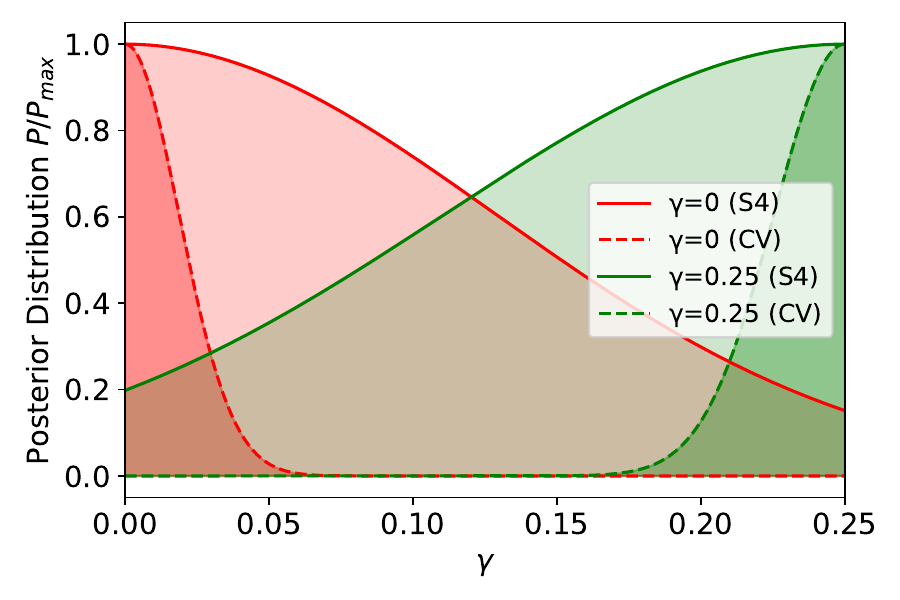}
    \caption{\color{black}Comparison of the (normalized) posterior distributions of the parameter $\c$ for the modified DW distribution.}
    \label{fig:gamma_comp}
\end{figure}
\begin{figure*}[htbp]
    \centering
    \includegraphics[width=0.9\linewidth]{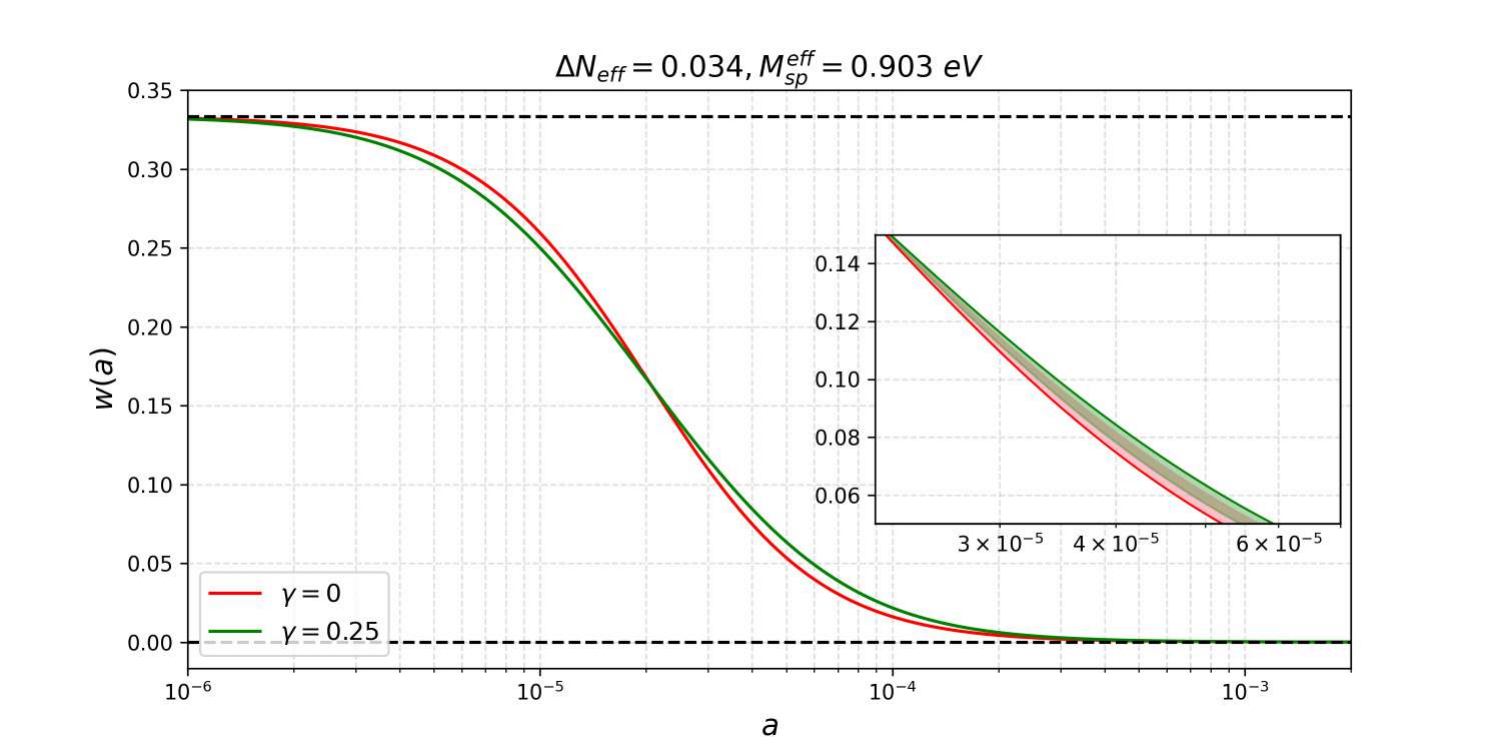}
    \caption{$w$ vs $a$ plot for the modified distribution (\ref{finalf}) with $\c=0$(DW) and $\c=0.25$. Shaded regions in the inset corresponds to the one-sided uncertainties on $w$ derived from the uncertainties on $\c$.}
    \label{fig:eos_moments}
\end{figure*}
{\color{black}Fig.~\ref{fig:gamma_comp} shows the truncated posterior distributions, marginalized only over the six $\La$CDM parameters. We quantify the distinguishability between the two cases by computing the overlap fraction between the (un-normalized) posterior probability distributions}. {\color{black}The analysis reveals a $\sim45\%$ probability mass overlap reaffirming that CMB-S4 data would not be able to conclusively distinguish between these models. The expected sensitivity of SO-LAT being less than that of the CMB-S4 survey, this then implies that the Simons Observatory would also not be able to distinguish between these models.}

The physical quantity related to the second moment of the LiMR distribution function is the "pressure" of the LiMRs. We define the quantity $f_P$, the \textit{fractional pressure} of the LiMRs, as ($g_s''$ is a degeneracy factor)
\begin{align}
    f_P\equiv \frac{P_\rm{sp}}{P_\n}=\frac{g_s''}{15/16}\underbrace{\l[\int dp~p^4\hat f(p)\r]}_{Q^{(2)}}/\l[\frac{45\zeta(5)}{2}(T_\n^\rm{id})^5\r].
\end{align} While the fiducial values of all the $\La$CDM parameters as well as that of $\Delta N_\rm{eff}$ and $M_\rm{sp}^\rm{eff}$ are the same, the fiducial values of $f_P$ differ between the cases $\c=0$ and $\c=0.25$. This is a consequence of differing second moments as mentioned above. {\color{black}The fiducial values of $f_P$ and the projected uncertainties (one-sided) are given in the following table:}
\begin{table}[htbp]
    \centering
    \begin{tabular}{|c|c|c|}
    \hline
       Experiment  &  $f_P[\c=0]$  &  $f_P[\c=0.25]$  \\\hline
        CMB-S4 & $0.180^{+0.260}$ & $0.265_{-0.301}$ \\\hline
        CV-limited & $0.180^{+0.050}$ & $0.265_{-0.094}$\\\hline
    \end{tabular}
    \caption{\color{black}Fiducial values and predicted constraints for the quantity $f_P$.}
    \label{tab:compgamma_fP}
\end{table}
{\color{black}where we have translated the uncertainties on $\c$ to those on the second moment by using $Q^{(2)}=Q^{(2)}_{DW}(=0.79325)+1.4865\c$.}
{\color{black} The fact that $f_P+\sigma(f_P)$ for $\c=0$ exceeds $0.265$ or $f_P+\sigma(f_P)$ becomes negative for $\c=0.25$ (in the first row above), both of which are unphysical, is partly due to the non-Gaussian nature of the posteriors and mostly because of the inadequate sensitivity of the CMB-S4 experiment.}
In Fig. \ref{fig:eos_moments}, we plot the equation of state parameter $w$ as a function of the scale factor for the modified distribution, for $\chi=0.034$ and $m_\rm{sp}=26.43~\rm{eV}$, with $\c=0$ and $\c=0.25$. The difference between the two curves arises from the difference in the pressures, or equivalently $f_P$, and not from $\rho_\rm{sp}$, which is the same for both cases. The inset plot zooms in on a region of the curves and shows the $1\sigma$ uncertainties on $w$, derived from the uncertainties on $\c$. The overlap between the pink and green shaded regions again indicates that CMB-S4 data would not be able to rule out one model in favor of the other.

\begin{figure}[htbp]
    \centering
    \includegraphics[width=\linewidth]{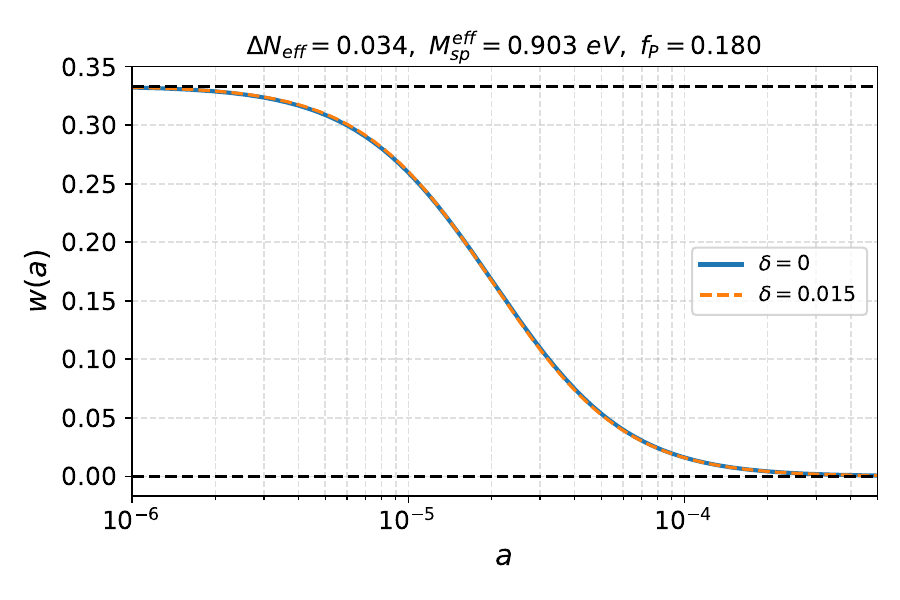}
    \caption{Evolution of $w$ with $a$ for the DW and the modified distribution when the first three moments are matched and the fourth moment is made to differ by the maximum possible extent. The exact overlap between the curves indicates the indistinguishability of the two models.}
    \label{fig:delta}
\end{figure}

If we include a $\d y^3$ term in the expansion (\ref{fq}) and match the first three moments, we find that the new parameter $\d$ can lie in the range $[0,0.015]$, where the upper bound {\color{black}on the free parameter $\d$} is an order of magnitude lesser than {\color{black}that on the free parameter $\c$ of} the previous case. This suggests that matching the first three moments essentially fixes the form of the modified distribution function with little room for deviation from the DW case. Similar to Fig.~\ref{fig:eos_moments}, in Fig.~\ref{fig:delta}, we plot the evolution of the equation of state parameter $w$ as a function of the scale factor $a$ for the new modified distribution with $\d=0$ and $\d=0.015$. We find that the curves corresponding to $\d=0$ and $\d=0.015$ overlap. Thus we conclude that no future CMB experiment would be able to distinguish between two LiMR models which agree on the first three moments of the corresponding LiMR distribution functions. This is likely to generalize to the case of matching even higher moments. Note that this conclusion may not extend to non-linear scales, a matter that is currently under investigation.

\section{Conclusion}\label{sec:conclusion}
The presence of LiMRs affects the evolution of the Universe. When relativistic, they behave as dark radiation and contribute to the relativistic energy density of the Universe,  whereas when non-relativistic, they behave as dark matter. The upcoming CMB experiments are expected to measure various cosmological parameters with unprecedented precision. In the context of light relics, $N_\rm{eff}$ is an important parameter. If the measured value of $N_\rm{eff}$ exceeds the SM prediction of $3.044$, it would indicate the presence of dark sector light relics (or the need to modify the thermal history of the Universe). {\color{black}SO is expected to achieve a sensitivity of $\sigma(\Delta N_\rm{eff})=0.045$~\cite{SimonsObservatory:2025wwn} and conservative configurations of the CMB-S4 experiment could reach $\sigma(N_\rm{eff})\sim0.02-0.03$~\cite{abazajian2016cmbs4sciencebookedition}.} For LiMRs, another important parameter is the relic mass, or rather its contribution to the non-relativistic energy density of the Universe today, captured by the parameter $M_\rm{sp}^\rm{eff}$. 
Another important aspect
of particles contributing to $N_{\rm{eff}}$ is their momentum distribution which is often tied to interesting aspects of the 
physics of the early Universe and particle physics such as production mechanism and interaction rates. Therefore, extracting the relation between distribution functions and observables is an important question.

In this work, we performed Fisher analyses to check how much {\color{black}future CMB data} would be able to constrain these parameters for \textit{non-thermal} LiMR models. We primarily concerned ourselves with a model of inflaton/moduli decay giving rise to LiMRs of mass $m_\rm{sp}$, with a branching ratio $B_\rm{sp}$. The LiMRs so produced follow a non-thermal distribution, unlike thermalized LiMRs which follow Fermi-Dirac or Bose-Einstein distribution, depending on the statistics~\cite{Baumann_2018,Xu_2022}. We also considered the Dodelson-Widrow distribution throughout. The main conclusions of our work are as follows:
\begin{itemize}
    \item {\color{black}Our Fisher analysis predicts a} tighter constraint on $\Delta N_\rm{eff}~(\sim 10^{-3})$ for a LiMR which becomes non-relativistic well before recombination, around the epoch of matter-radiation equality. However, the {\color{black}projected uncertainty} so obtained might be misleading owing to the fact that the LiMR does not contribute to $\Delta N_\rm{eff}$ at recombination, which is a measure for additional ``relativistic" degrees of freedom. The {\color{black}smaller projected uncertainty} (further enhanced by lensing) is owing to the fact that changes in $\w_m(=\w_b+\w_\rm{cdm})$ or $\w_\rm{sp}$ cannot be replicated by changes in $\Delta N_\rm{eff}$ which is zero by recombination. For a LiMR which transitions in the matter dominated era and becomes completely non-relativistic just after recombination, we obtained a {\color{black}less precise forecast} on $\Delta N_\rm{eff}~(\sim 10^{-2})$ in line with the expected uncertainty for light relics~\cite{abazajian2016cmbs4sciencebookedition,SimonsObservatory:2025wwn}.

    \item Our analysis is section~\ref{sec:discussion} (see Fig.~\ref{fig:fisher_ellipses}) also revealed interesting correlations between model parameters and observables. In particular, we obtained negative correlation between the phenomenological parameters for set I but positive correlation for set II. These can be relevant in 
    endeavors to connect cosmology with particle physics model building.

    \item We compared our results for the non-thermal model with those of the Dodelson-Widrow model, and found that CMB-S4 data would yield similar {\color{black}constraints} on the parameters for these models once the phenomenological parameters $\Delta N_\rm{eff}$ and $M_\rm{sp}^\rm{eff}$ (related to the first two moments of the distribution function) are matched.

    \item We also compared the Dodelson-Widrow distribution with a modified distribution (which can be considered as a perturbed Dodelson-Widrow distribution), artificially constructed so as to match the first two moments but differing to the maximum possible extent in the higher moments. We deduced that CMB-S4 data would be unable to tell even these two distributions apart. We believe this generalizes to any two distribution functions with matching first two moments.

\end{itemize}
LiMRs not only affect the CMB primary and lensing power spectra but also affect the matter power spectrum due to their free streaming/clustering effects. In fact, the effects on the matter power spectrum feed back to the CMB lensing power spectrum. As such, future galaxy surveys such as LSST and Euclid (VRO) can also probe the impact of LiMRs, and one can perform Fisher analyses for such galaxy surveys. Combining CMB and LSS information is crucial for obtaining optimal constraints on LiMRs, particularly for those relics which are not fully relativistic at recombination, as has been observed in Ref.~\cite{Xu_2022}, for example. This we keep for future work. Moreover, going beyond linear theory to the non-linear matter power spectrum unlocks new information about LiMRs --- chiefly through the distinctive, scale-dependent damping they induce, which cannot be mimicked by simply retuning the $\La$CDM parameters~\cite{Banerjee:2022era}. This indicates that degeneracies that exist on large (linear) scales can be broken by using signatures on small (non-linear) scales. In Ref.~\cite{Banerjee:2022era}, the authors also found that non-linear effects can reveal differences between LiMR models that appear identical at the linear level. Remarkably, these distinctions arise from differences in the momentum distribution functions of LiMRs. Since these distributions are intimately connected to the production mechanisms of non-thermal relics, the resulting non-linear signatures could provide valuable insight into early Universe physics—insight that linear analyses, including the present study, are inherently insensitive to.  

\begin{acknowledgments}
{\color{black}The authors thank the anonymous referee whose suggestions helped us improve this paper significantly.} RKS thanks the Alexander von Humboldt Foundation for their support. AB acknowledges support from the Science and Engineering Research Board (SERB) India
via the Startup Research Grant SRG/2023/000378.
\end{acknowledgments}

\appendix
\section{INFLATON/MODULI DECAY MODEL}\label{appendix:A}
In this appendix, we discuss briefly the production mechanism leading to the form of the non-thermal distribution function (\ref{fqNT}) for the LiMRs. Further details can be found in Ref.~\cite{Bhattacharya:2020zap}. We start with a matter-dominated Universe at a (dimensionless) 'initial time'($\theta\equiv t/\tau=0$) and evolve the Universe using the following equations up to a time $\theta^*$ which is large enough to ensure that almost all the $\phi$ particles have decayed\footnote{We choose $\theta^*=15$ in practice.}:
\begin{align}\label{4}
    \dot\rho_m+3H\rho_m=-\frac{\rho_m}{\tau},
\end{align}
\begin{align}\label{5}
    \dot\rho_r+4H\rho_r=+\frac{\rho_m}{\tau},
\end{align}
and
\begin{align}\label{6}
    H=\frac{\dot a}{a}=\sqrt{\frac{\rho_m+\rho_r}{3M_\rm{Pl}^2}},
\end{align}
where $\rho_m=\rho_\phi$ denotes the energy density in matter and $\rho_r=\rho_\rm{sp}+\rho_\rm{SM}$ is the energy density in radiation. As a result of decays, the comoving inflaton/modulus number density falls off as $N(t)=N(0)e^{-t/\tau}$ where
\begin{align}\label{7}
    N(0)\equiv\frac{\rho_m(0)}{m_\phi}=\frac{4\a M_\rm{Pl}^2}{3\tau^2m_\phi}\quad(\a\gg1).
\end{align}
Combining this with the fact that the energy of a particle produced at $t=t_d$ will evolve according to (with $\hat E=E(t_d)$)
\begin{align}\label{8}
    E(t)=\hat E\l(\frac{a(t_d)}{a(t)}\r)=\frac{m_\phi}{2}\l(\frac{a(t_d)}{a(t)}\r),
\end{align}
one can get the physical number density spectrum of the particles,
\begin{align}\label{A6}
    dN_t&=\frac{1}{a^3(t)}\frac{2B_\rm{sp}}{\tau}N(0)e^{-t_d/\tau}dt_d\nn\\
    &=\frac{2B_\rm{sp}}{\hat s^3(\theta)\tau}N(0)e^{-\hat s^{-1}(y)}\frac{dE}{E\hat H(\hat s^{-1}(y))},
\end{align}
{\color{black}where we have introduced the variable $y\equiv E\hat s(\theta)/\hat E$. The number density spectrum is non-vanishing when $y$ varies between between $1$ (corresponding to decays at the initial time) and $\hat s(\theta)$ (corresponding to decays at $\theta$). Using isotropy we can write}
\begin{align}
    dN_t=\tilde n(E)dE=\frac{\tilde n_t(|\bs p|)}{4\pi|\bs p|^2}d^3p\equiv n_t(\bs p)d^3p,
\end{align}
{\color{black}where $n_t(\bs p)=n_{t^*}\l(\frac{a(t)}{a(t_*)}\bs p\r)$ since the sterile LiMRs free stream after production. The bounds on $y$ translate to those on $|\bs p|$; $n_{t_0}(\bs p)$ is non-vanishing if}
\begin{align}\label{A8}
    \frac{\hat E}{a(t_0)}<|\bs p|<\frac{\hat Ea(t^*)}{a(t_0)}.
\end{align}
One hence arrives at the following late-time momentum distribution function by red-shifting the energy of the particles from their time of production
\begin{align}
    f(\bs q)=\frac{32}{\pi\hat E^3}\l(\frac{N(0)B_\rm{sp}}{\hat s^3(\theta^*)}\r)\frac{\exp{(-\hat s^{-1}(y))}}{|\bs q|^3\hat H(\hat s^{-1}(y))},
\end{align}
where $y=(|\bs q|/4)\hat s(\theta^*)$ and $|\bs q|$ is constrained so that
\begin{align}\label{10}
    \frac{4}{\hat s(\theta^*)}<|\bs q|<4.
\end{align}
The argument of the distribution function is $|\bs q|\equiv |\bs p|/T_\rm{ncdm,0}$ where~\cite{Bhattacharya:2020zap}
\begin{align}\label{Tncdm0}
    T_\rm{ncdm,0}=0.418\l(\frac{m_\phi^2\tau}{M_\rm{Pl}}\r)^{1/2}\frac{T_\rm{cmb}}{(1-B_\rm{sp})^{1/4}}\equiv\zeta T_\rm{cmb}
\end{align}
is the temperature of the non-cold-dark matter species today, expressed in terms of the CMB temperature.

{\color{black}The expressions for the phenomenological parameters are obtained as follows. $\Delta N_\rm{eff}$ is given by}
\begin{align}
        \Delta N_\rm{eff}&\equiv\frac{\rho_\rm{sp}}{\rho_\n}\bigg\rvert_\rm{\n,dec.}=\frac{\rho_\rm{SM}}{\rho_\n}\bigg\rvert_\rm{\n,dec.}\times \frac{\rho_\rm{sp}}{\rho_\rm{SM}}\bigg\rvert_\rm{\n,dec.}\nn\\
        &=\frac{g_\rm{*SM}}{g_{*\n}}\bigg\rvert_\rm{\n,dec.}\times \l(\frac{g_*(T(t_\n))}{g_*(T(t^*))}\r)^{1/3}\frac{B_\rm{sp}}{1-B_\rm{sp}}\nn\\
        &=\frac{10.75}{1.75}\l(\frac{g_*(T(t_\n))}{g_*(T(t^*))}\r)^{1/3}\frac{B_\rm{sp}}{1-B_\rm{sp}},
    \end{align}
{\color{black}which is typical for models of inflaton/modulus decay to dark radiation particles (see, for example, Refs.~\cite{Cicoli_2013,1304.1804}). Now $\w_\rm{sp}$ (and hence $M_\rm{sp}^\rm{eff}$) can be obtained by taking the ratio of the number density of LiMRs at $t^*$, $n_\rm{sp}(t^*)=2B_\rm{sp}N(0)/a^3(t^*)$, and the entropy density at $t^*$, $s(t^*)=\frac{4}{3}\rho_\rm{SM}^3(t^*)\l(\frac{\pi^2}{30}g_*(t^*)\r)^{1/4}$, and using the fact that this ratio is conserved after $t^*$. The present day LiMR number density $n_\rm{sp,0}=\frac{n_\rm{sp}(t^*)}{s(t^*)}s_0$ can in turn be expressed in terms of $n_{\n,0}$, which upon substituting in}
\begin{align}
    \w_\rm{sp}=m_\rm{sp}n_\rm{sp,0}\l(\frac{h^2}{\rho_c^0}\r),
\end{align}
{\color{black}and multiplying the resulting expression by $94.05~\rm{eV}$ yields Eq. (\ref{Meff}).}

\section{NUMERICAL DERIVATIVES}\label{appendix:B}
The numerical differentiation was done using the three-point central difference formula with equal left and right step sizes (for two-sided derivatives)
\begin{align}
    \frac{\pa C_\ell(\theta_0)}{\pa \theta}=\frac{C_\ell(\theta_0+\Delta s)-C_\ell(\theta_0-\Delta s)}{2\Delta s},
\end{align}
or the two-point forward/backward difference formula (for one-sided derivatives)
\begin{equation}
\begin{split}
    \frac{\pa C_\ell(\theta_0)}{\pa \theta}\Bigg\rvert_{+}&=\frac{C_\ell(\theta_0+\Delta s)-C_\ell(\theta_0)}{\Delta s},\\
    \frac{\pa C_\ell(\theta_0)}{\pa \theta}\Bigg\rvert_{-}&=\frac{C_\ell(\theta_0)-C_\ell(\theta_0-\Delta s)}{\Delta s}.
\end{split}
\end{equation}
Here $\theta_0$ is the fiducial value of the parameter $\theta$ under consideration and $\Delta s$ is the step size. We have not accounted for the dependence of $N_\ell^{\phi\phi}$ on the parameters (through the $C_\ell$ values) while computing the derivatives---$N_\ell^{\phi\phi}$'s were kept fixed at their fiducial values.

\begin{table}[htbp]
    \centering
    \begin{tabular}{|c|c|c|c|c|}
        \hline
        Parameter & \multicolumn{2}{c|}{Fid. val. Set I} & \multicolumn{2}{c|}{Fid. val. Set II} \\
        \cline{2-5}
                  & fiducial & step & fiducial & step \\
        \hline
        $\omega_b$     & 0.02247        & 1.6\% & 0.02242 & 6\% \\
        $\omega_{\text{cdm}}$ & 0.111  & 4\% & 0.120 & 4\% \\
        $h$            & 0.6804         & 3\% & 0.6711 & 3\% \\
        $10^9 A_s$     & 2.099          & 1\% & 2.110 & 1\% \\
        $n_s$          & 0.9661         & 1\% & 0.9652 & 1\% \\
        $\tau_{\text{reio}}$ & 0.0536   & 4\% & 0.0560 & 6.4\% \\\hline   
        $B_{\text{sp}}$ & 0.0118        & 1.6\% &  0.0332 & 5\% \\
        $m_{\text{sp}}$~[eV] & 38.62    &  8\%   &   6.2  & 5\% \\
        \hline
    \end{tabular}
    \caption{\color{black}Fiducial values and step sizes (reported as percentage of fiducial value) for the six $\La$CDM parameters and the NT model parameters used in the Fisher analysis for the two sets of fiducial values considered in the text.}
    \label{tab:stepsizes}
\end{table}

\begin{figure*}[htbp]
    \centering
    \begin{subfigure}{\textwidth}
        \centering
        \includegraphics[width=\textwidth]{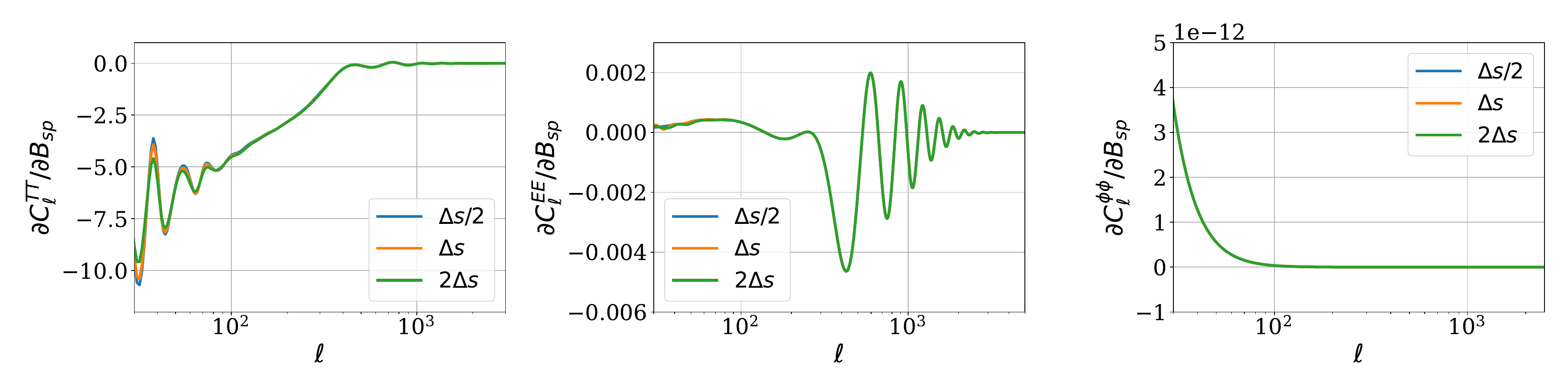}
        \label{fig:sub11}
    \end{subfigure}
    \hfill
    \begin{subfigure}{\textwidth}
        \centering
        \includegraphics[width=\textwidth]{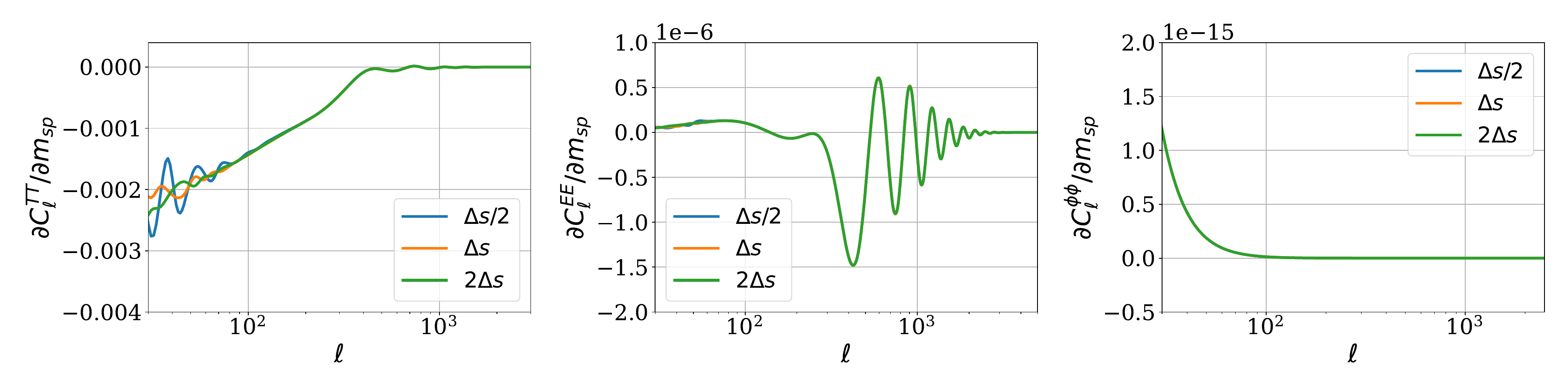}
        \label{fig:sub22}
    \end{subfigure}
    \caption{\color{black}Convergence plots for the derivatives of the CMB primary and lensing power spectra with respect to $B_\rm{sp}$ (top) and $m_\rm{sp}$ (bottom) for the NT model (set I). }
    \label{fig:dCl_conv}
\end{figure*}

\begin{figure*}[htbp]
    \centering
    \begin{subfigure}{\textwidth}
        \centering
        \includegraphics[width=\textwidth]{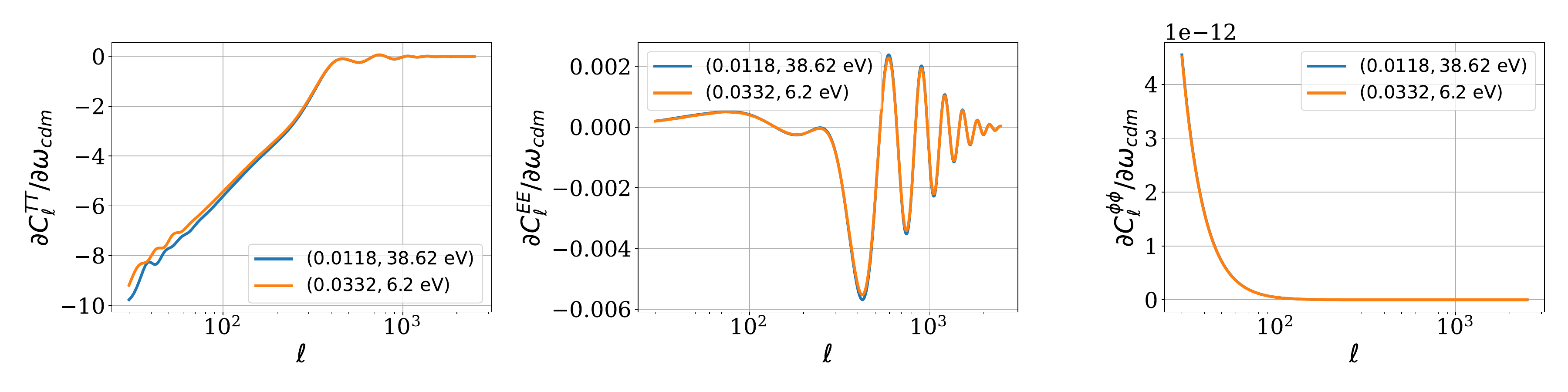}
        \label{fig:sub1}
    \end{subfigure}
    \hfill
    \begin{subfigure}{\textwidth}
        \centering
        \includegraphics[width=\textwidth]{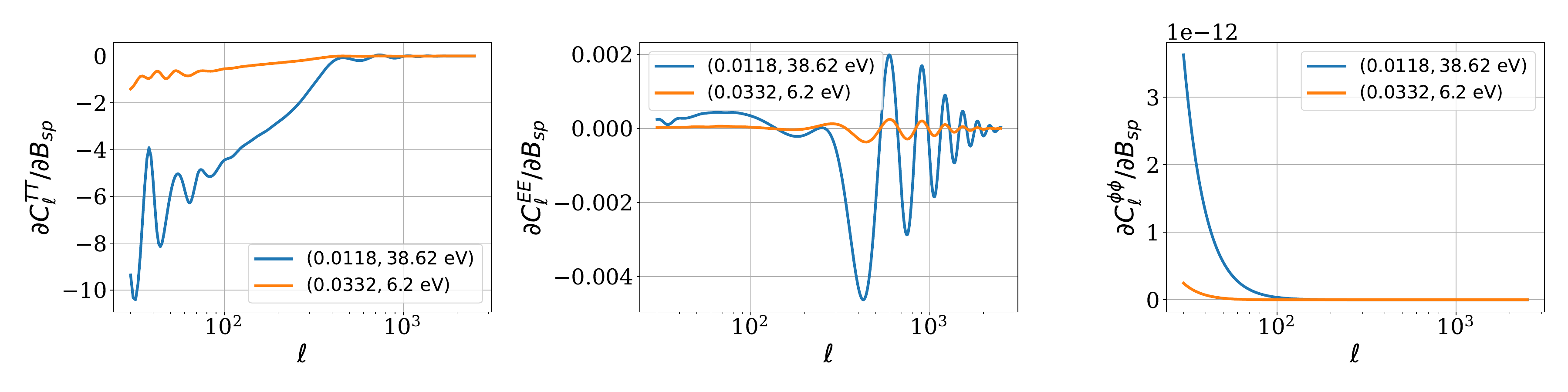}
        \label{fig:sub2}
    \end{subfigure}
    \caption{\color{black}Comparative plots for the derivatives of the CMB primary and lensing power spectra with respect to $\w_\rm{cdm}$ (top) and $B_\rm{sp}$ (bottom) for the NT model for the two sets of fiducial values.}
    \label{fig:dCl}
\end{figure*}

Numerical derivatives play a crucial role in Fisher analysis, and in the computation of numerical derivatives, the choice of step sizes is important. Too large a step size misses fine features and introduces systematic errors, while derivatives are prone to numerical noise and round off errors if the step size is too small. {\color{black} To check for the convergence of the derivatives, we plotted the derivatives with step sizes $\Delta s$ (given in Table \ref{tab:stepsizes} for the NT model), $\Delta s/2$ and $2\Delta s$. We first visually inspected the derivatives to check that the curves overlap across multipoles, and then computed two diagnostic quantities,
\begin{align}
    \eps_1&=\frac{\sum_\ell|C_\ell'(\Delta s)-C_\ell'(\Delta s/2)|}{\sum_\ell|C_\ell'(\Delta s)|},\\
    \quad\eps_2&=\frac{\sum_\ell|C_\ell'(2\Delta s)-C_\ell'(\Delta s)|}{\sum_\ell|C_\ell'(\Delta s)|},
\end{align}
and checked that $\eps_{1,2}\lesssim10^{-2}$. In Fig.~\ref{fig:dCl_conv}, we show the convergence of the derivatives of $C_\ell^{TT},C_\ell^{EE},$ and $C_\ell^{\phi\phi}$ with respect to the model-specific parameters $B_\rm{sp}$ and $m_\rm{sp}$ for the NT model (set I). Although small oscillatory differences are visible at low multipoles ($\ell\lesssim100$) for some parameters (like $m_\rm{sp}$) we verified that excluding $\ell<100$ has a negligible impact on our results pertaining to the phenomenological parameters (and for most cases, we found $\eps_{1,2}\sim10^{-3}$ when we excluded the $\ell<100$ region).}
In Fig. \ref{fig:dCl}, we compare the numerical derivatives of $C_\ell^{TT}$, $C_\ell^{EE}$, and $C_\ell^{\phi\phi}$ with respect to two parameters, $\w_\rm{cdm}$ and $B_\rm{sp}$, for both sets of fiducial values. We see that the derivatives with respect to $\w_\rm{cdm}$ are nearly identical for both parameter sets. This observation holds for all six $\La$CDM parameters, which indicates that the direct effect of these parameters on the CMB spectra is dominant and largely independent of the specific LiMR properties. In contrast, the derivatives with respect to $B_\rm{sp}$ (and also $m_\rm{sp}$) are significantly different between the two parameter sets, which highlights the sensitivity of the CMB power spectra to the LiMR properties, like its branching ratio and its mass. Computationally, this is the primary source of the difference in the uncertainty values in $\Delta N_\rm{eff}$ and $M_\rm{sp}^\rm{eff}$ between the two sets of fiducial values for the NT model. Similar arguments hold for the DW model.

\section{\color{black}EXPLANATION OF PARAMETER CORRELATIONS}\label{appendix:C}
First, let us consider the heavier LiMR of set I. Increasing the LiMR's contribution to $\Delta N_\rm{eff}$ (by increasing $B_\rm{sp}$ or $\chi$) only adds to the matter content (through $M_\rm{sp}^\rm{eff}$) around matter-radiation equality as well as at recombination. This has the following consequences: (a) a greater decrease in the angular diameter distance at recombination $D_{A*}$ as compared to the sound horizon at recombination $r_{s*}$, thereby increasing $\theta_*=r_{s*}/D_{A*}$, and (b) shifting $z_\rm{eq}$ to an earlier time. Both $\theta_*$ and $z_\rm{eq}$ are well constrained by observations~\cite{Planck:2018vyg}. Because late-time LiMR abundance scales linearly with the LiMR mass $m_\rm{sp}$ for a fixed comoving number density ($\propto B_\rm{sp}$ or $\chi$), decreasing $M_\rm{sp}^\rm{eff}$ (via $m_\rm{sp}$) removes the extra LiMR matter introduced, leaving the early relativistic energy density essentially unchanged. The increased radiation content in the early universe is offset by increasing $\w_m$ which restores $1+z_\rm{eq}\propto(\w_m+\w_\rm{sp})/\w_r$ to its fiducial value. Although changing only $M_\rm{sp}^\rm{eff}$ or $\w_m$ appropriately could, in principle, keep $z_\rm{eq}$ unchanged, such substitutions are disfavored since LiMR cannot mimic CDM at early times (for example, CDM tends to suppress early ISW effect while LiMR tends to boost it). Even after compensating the $\Delta N_\rm{eff}$-induced changes through $M_\rm{sp}^\rm{eff}$ and $\w_m$, the residual fractional shifts in $D_{A*}$ and $r_{s*}$ do not cancel. This leads to a decrease in $h$ (which does not affect $z_\rm{eq}$) in order to keep $\theta_*$ (almost)unchanged. Note that this is contrary to what happens for standard (massless) neutrinos---$h$ and $N_\rm{eff}$ are positively correlated~\cite{Planck:2018vyg}. Thus we have explained the negative correlations between $(\Delta N_\rm{eff}-M_\rm{sp}^\rm{eff})$ and $(\Delta N_\rm{eff}-h)$ and the positive correlation between $(\Delta N_\rm{eff}-\w_m)$.\\
{For the lighter particle of set II, the situation is somewhat different. Now the epoch of matter-radiation equality is set by
\begin{align*}
    \rho_ra_\rm{eq}^{-4}\rm+f_\rm{eq}\rho_\rm{sp}a_\rm{eq}^{-4}=\rho_ma_\rm{eq}^{-3}+(1-f_\rm{eq})\rho_\rm{sp}a_\rm{eq}^{-3},
\end{align*}
where $f_\rm{eq}=3w_\rm{sp}(a_\rm{eq})\approx0.15$. Increasing $\Delta N_\rm{eff}$ again leads to a decrease in $m_\rm{sp}$, but nevertheless leaves $M_\rm{sp}^\rm{eff}$ larger than its fiducial value because $f_\rm{eq}$ is non-zero. Hence the (weakly)positive correlation between $(\Delta N_\rm{eff}-M_\rm{sp}^\rm{eff})$ (degeneracy direction mostly along $\Delta N_\rm{eff}$). Although a further decrease in $m_\rm{sp}$ so as to offset the increase in $M_\rm{sp}^\rm{eff}$ is allowed in principle, such a change is disfavored due to the following reason. Further lowering $m_\rm{sp}$ would extend the LiMR's free-streaming suppression into scales probed by CMB lensing in a way that cannot be fully compensated by changes in $\w_m$.\footnote{For the same argument to hold for the LiMR of set I, its mass would need to decrease approximately by an order of magnitude.} Similar arguments as used for set I also explain the positive correlation between $(\Delta N_\rm{eff}-\w_m)$. Without lensing, we found a weak positive correlation between $(\Delta N_\rm{eff}-h)$. Also the change in the sign of correlation between $M_\rm{sp}^\rm{eff}$ and $\w_m$ (flips from slightly negative to positive) is due to the inclusion of the lensing effects, \ie not influenced by pre-recombination era physics. More $\w_m$ enhances lensing strength, compensating for the suppression due to the relic's free streaming. But this again decreases $D_{A*}$ more than $r_{s*}$; $h$ then decreases to restore $\theta_*$, \ie lensing indirectly induces the negative correlation between $(\Delta N_\rm{eff}-h)$.
\begin{figure}[t]
    \centering
    \begin{subfigure}{0.48\textwidth}
        \centering
        \includegraphics[width=\textwidth]{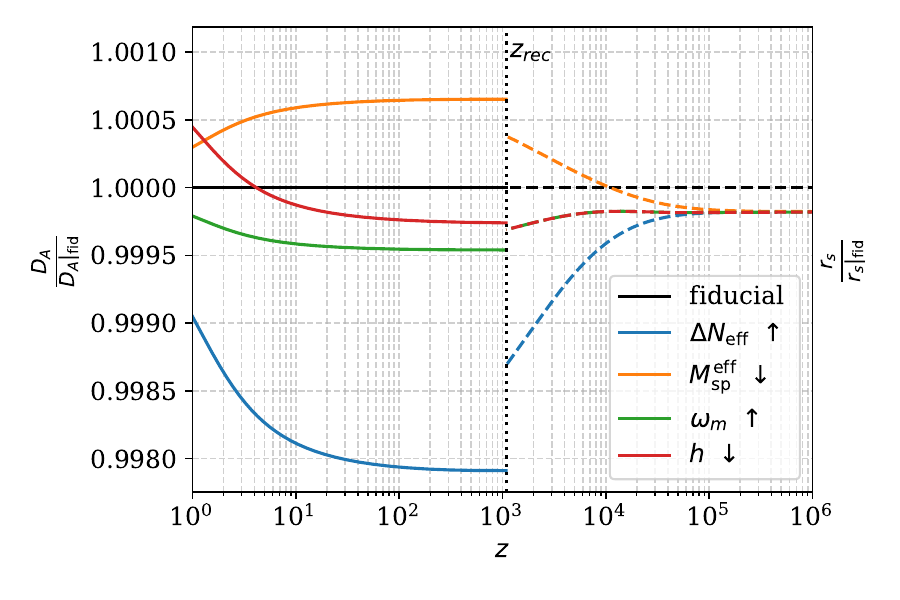}
        \caption{Effect of increasing/decreasing the parameter values (with respect to the fiducial values) on $\theta_*=r_{s*}/D_{A*}$ for set I.}
    \end{subfigure}
    \hfill
    \begin{subfigure}{0.48\textwidth}
        \centering
        \includegraphics[width=\textwidth]{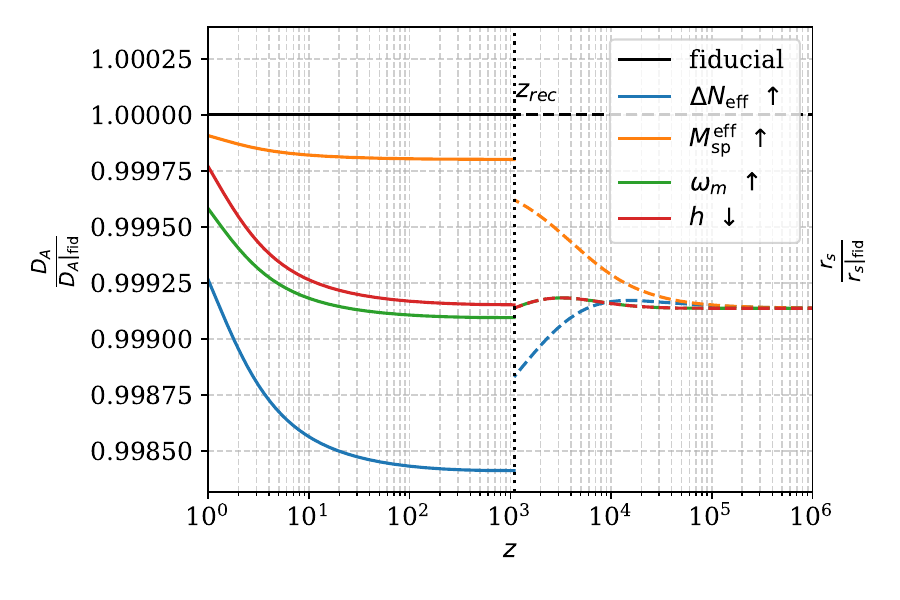}
        \caption{Same as (a) for set II.}
    \end{subfigure}
    \caption{\color{black}Quantitative validation of the degeneracies between the set of parameters $\{\Delta N_\rm{eff},M_\rm{sp}^\rm{eff},h,\w_m\}$ as they relate to $\theta_*$; $h$ affects $\theta_*$ only through its contribution to $D_{A*}$.}
    \label{fig:disc1}
\end{figure}
Since a CV-limited experiment is expected to reveal the intrinsic physical correlations between different parameters (particularly those whose degeneracies are not dominated by low-$\ell$, CV-dominated regime), we have tried to explain the correlations seen in Fig.~\ref{fig:CVlensunlens}.\footnote{In real experiments, low signal-to-noise ratio at higher multipoles can re-weight contributions and occasionally flip the apparent sign of a correlation.} In Table~\ref{tab:disc1} and Fig.~\ref{fig:disc1} we provide the quantitative validation of our qualitative arguments. In generating the table and the plots, we have considered the NT model (for concreteness) and increased $B_\rm{sp}$ by $2\sigma$ followed by increasing/decreasing $\{\w_b,\w_\rm{cdm},m_\rm{sp},h\}$ by $\leq2\sigma$. 
\begin{table}[b]
\centering
\setlength{\tabcolsep}{3pt} 
\begin{tabular}{|c|c|c|c|c|c|}
\hline
$\w_m$ & $h$ & $\Delta N_\rm{eff}$ & $M_\rm{sp}^\rm{eff}~[\rm{eV}]$ & $z_\rm{eq}$ &\% change\\
\hline
\multicolumn{6}{|c|}{Set I} \\ 
\hline
$\textit{0.13347}$ & $\textit{0.6804}$ & $\textit{0.034}$ & $\textit{0.903}$ & $\textit{3421.85}$ & - \\\hline
$0.13347$ & $0.6804$ & $\textbf{0.037}$ & $0.972$ & $3439.65$ & $+0.25$\\\hline
$0.13347$ & $0.6804$ & $\textbf{0.037}$ & $\textbf{0.881}$ & $3415.82$ & $-0.44$\\\hline
$\textbf{0.13387}$ & $0.6804$ & $\textbf{0.037}$ & $\textbf{0.881}$ & $3425.36$ & $+0.10$ \\\hline
$\textbf{0.13387}$ & $\textbf{0.6797}$ & $\textbf{0.037}$ & $\textbf{0.881}$ & $3425.36$ & $+0.10$\\\hline
\multicolumn{6}{|c|}{Set II} \\ 
\hline
$\textit{0.14242}$ & $\textit{0.6711}$ & $\textit{0.100}$ & $\textit{0.415}$ & $\textit{3478.92}$ & - \\\hline
$0.14242$ & $0.6711$ & $\textbf{0.113}$ & $0.449$ & $3488.44$ & $+0.27$ \\\hline
$0.14242$ & $0.6711$ & $\textbf{0.113}$ & $\textbf{0.421}$ & $3474.04$ & $-0.14$ \\\hline
$\textbf{0.14268}$ & $0.6711$ & $\textbf{0.113}$ & $\textbf{0.421}$ & $3480.14$ & $+0.03$\\\hline
$\textbf{0.14268}$ & $\textbf{0.6709}$ & $\textbf{0.113}$ & $\textbf{0.421}$ & $3480.14$ & $+0.03$
\\\hline
\end{tabular}
\caption{\color{black}Quantitative validation of the degeneracies between the set of parameters $\{\Delta N_\rm{eff},M_\rm{sp}^\rm{eff},\w_m=\w_b+\w_\rm{cdm}\}$ as they relate to $z_\rm{eq}$ (which is unaffected by changes in $h$). Fiducial values are shown in italics, and changes are shown in bold font.}
\label{tab:disc1}
\end{table}

\begin{figure*}[htbp]
    \centering
    \begin{subfigure}[t]{0.48\textwidth}
        \centering
        \includegraphics[width=\textwidth]{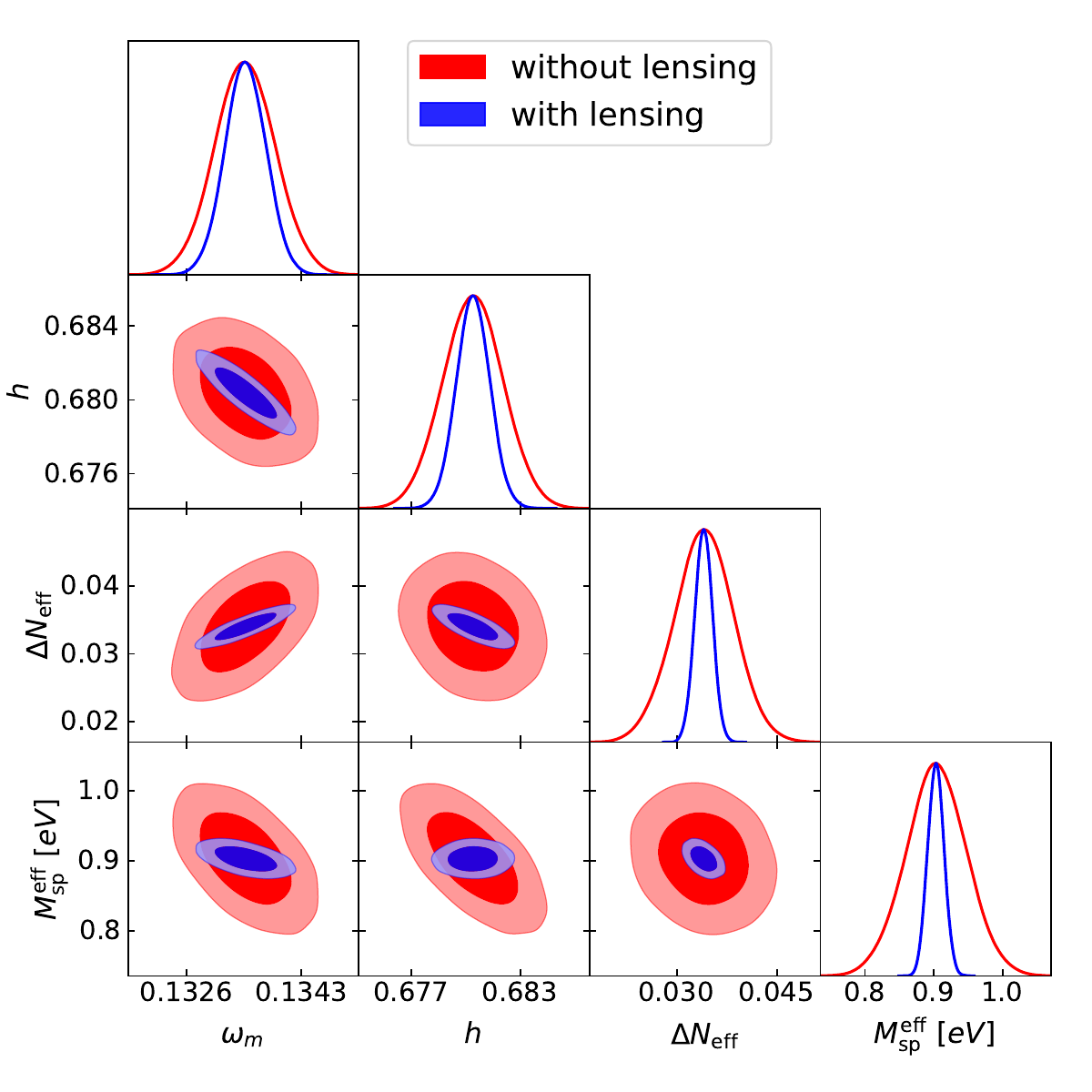}
        \caption{Fisher plots for the NT model, with and without lensing, for fiducial values set I.}
    \label{fig:nt2small}
    \end{subfigure}
    \hfill
    \begin{subfigure}[t]{0.48\textwidth}
        \centering
        \includegraphics[width=\textwidth]{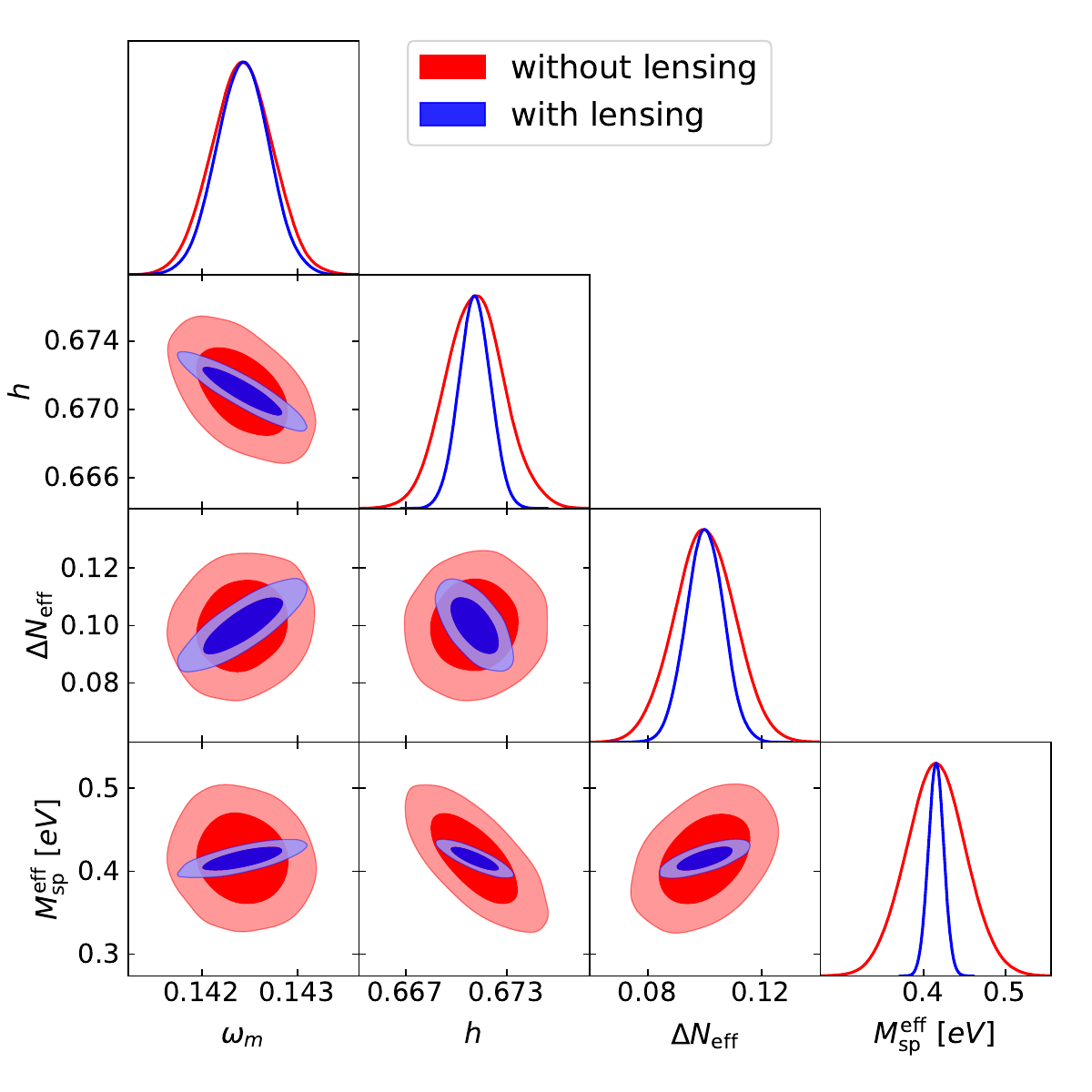}
        \caption{Same as (a) with fiducial values set II.}
    \label{fig:nt4small}
    \end{subfigure}
    \caption{\color{black}The posterior distributions and Fisher ellipses for the parameters $\{\omega_m=\w_b+\w_\rm{cdm},h,\Delta N_\rm{eff},M_\rm{sp}^\rm{eff}\}$ from the CV-limited experiment.}
    \label{fig:CVlensunlens}
\end{figure*}

\bibliographystyle{apsrev4-2}
\bibliography{main}
\end{document}